\documentclass[conference,10pt]{IEEEtran}
\usepackage{amsmath,amssymb,amsfonts}
\usepackage[T1]{fontenc}
\usepackage{graphicx}
\usepackage{textcomp}
\usepackage{bmpsize}
\usepackage{xcolor}
\usepackage{booktabs}
\usepackage{lipsum}
\usepackage[normalem]{ulem}
\usepackage{nicefrac}
\usepackage{cancel}
\usepackage{placeins}
\usepackage[backend=bibtex,style=ieee,natbib=true]{biblatex} 
\usepackage{siunitx}
\usepackage[colorlinks=true,urlcolor=black]{hyperref}

\def\BibTeX{{\rm B\kern-.05em{\sc i\kern-.025em b}\kern-.08em
    T\kern-.1667em\lower.7ex\hbox{E}\kern-.125emX}}
\def\BibTeX{{\rm B\kern-.05em{\sc i\kern-.025em b}\kern-.08em
    T\kern-.1667em\lower.7ex\hbox{E}\kern-.125emX}}

\newcommand{\mypar}[1]{\textbf{\emph{#1.}}}
\newcommand{\code}[1]{\texttt{#1}}

\newcommand{\expfunc}{\texttt{exp()}\xspace}
\newcommand{\expffunc}{\texttt{expf()}\xspace}

\newcommand{\expfmaxboundnb}{88.723}
\newcommand{\expfmaxbound}{$\expfmaxboundnb$\xspace}

\newcommand{\sigmoidfunc}{\texttt{sigmoid()}\xspace}
\newcommand{\tanhfunc}{\texttt{tanh()}\xspace}
\newcommand{\softplusfunc}{\texttt{softplus()}\xspace}
\newcommand{\softmaxfunc}{\texttt{softmax()}\xspace}
\newcommand{\gelufunc}{\texttt{GELU()}\xspace}
\newcommand{\swiglufunc}{\texttt{SwiGLU()}\xspace}

\newcommand{\relufunc}{\texttt{relu()}\xspace}

\newcommand{\pocI}{\texttt{M}\textsubscript{insurance}\xspace}
\newcommand{\pocII}{\texttt{M}\textsubscript{mult}\xspace}
\newcommand{\pocIII}{\texttt{M}\textsubscript{MNIST}\xspace}

\usepackage{enumitem}
\usepackage[capitalize]{cleveref}
\usepackage[nolist]{acronym}
\usepackage{pifont}
\usepackage{algorithm}
\usepackage{algpseudocode}
\newcommand{\cmark}{\ding{51}}%
\newcommand{\xmark}{\ding{55}}%
\algnewcommand\algorithmicforeach{\textbf{for each}}
\algdef{S}[FOR]{ForEach}[1]{\algorithmicforeach\ #1\ \algorithmicdo}

\newcommand{\ie}{\textit{i.e.},\ }
\newcommand{\eg}{e.g.,\ }
\newcommand{\na}{n.a.\ }

\newcommand{\cf}{cf.\ }
\newcommand{\sgxstep}{{SGX-Step}\xspace}
\newcommand{\tfulite}{Tensorflow Microlite\xspace}
\newcommand{\tflite}{Tensorflow lite\xspace}

\usepackage{tikz}
\usetikzlibrary{matrix,chains,positioning,decorations.pathreplacing,arrows}
\usepackage{xspace}
\definecolor{ForestGreen}{RGB}{34,139,34}
\usepackage{listings}
\AtBeginEnvironment{appendices}{\crefalias{section}{appendix}}

\crefname{lstlisting}{Listing}{Listings}
\Crefname{lstlisting}{Listing}{Listings}

\begin{document}

\begin{acronym}[ASCII]
 \setlength{\itemsep}{0.2em}
 \acro{3DES}{Triple DES}
 \acro{AES}{Advanced Encryption Standard}
 \acro{AMD}{Advanced Micro Devices}
 \acro{ANF}{Algebraic Normal Form}
 \acro{AoT}{Ahead of Time}
 \acro{API}{Application Programming Interface}
 \acro{ARM}{Advanced RISC Machines}
 \acro{ARX}{Addition, Rotate, XOR}
 \acro{ASK}{Amplitude-Shift Keying}
 \acro{ASIC}{Application Specific Integrated Circuit}
 \acro{BL}{Bootloader Enable}
 \acro{BOR}{Brown-Out Reset}
 \acro{BPSK}{Binary Phase Shift Keying}
 \acro{CBC}{Cipher Block Chaining}
 \acro{CBS}{Critical Bootloader Section}
 \acro{CGM}{Continuous Glucose Monitoring System}
 \acro{CMOS}{Complementary Metal Oxide Semiconductor}
 \acro{CNN}{Convolutional Neural Network}
 \acro{COPACOBANA}{Cost-Optimized Parallel Code Breaker and Analyzer}
 \acro{CPA}{Correlation Power Analysis}
 \acroplural{CPA}[CPAs]{Correlation Power Analyzes}
 \acro{CPU}{Central Processing Unit}
 \acro{CTR}{Counter \acroextra{(mode of operation)}}
 \acro{CRC}{Cyclic Redundancy Check}
 \acro{CRP}{Code Readout Protection}
 \acro{DES}{Data Encryption Standard}
 \acro{DC}{Direct Current}
 \acro{DDS}{Digital Direct Synthesis}
 \acro{DFT}{Discrete Fourier Transform}
 \acro{DMA}{Direct Memory Access}
 \acro{DNNL}{Deep Neural Network Library}
 \acro{DoS}{Denial-of-Service}
 \acro{DFA}{Differential Fault Analysis}
 \acro{DPA}{Differential Power Analysis}
 \acro{DRAM}{Dynamic Random-Access Memory}
 \acro{DRM}{Digital Rights Management}
 \acro{DSO}{Digital Storage Oscilloscope}
 \acro{DSP}{Digital Signal Processing}
 \acro{DST}{Digital Signature Transponder}
 \acro{DUT}{Device Under Test}
 \acroplural{DUT}[DUTs]{Devices Under Test}
 \acro{ECB}{Electronic Code Book}
 \acro{ECC}{Elliptic Curve Cryptography}
 \acro{ECU}{Electronic Control Unit}
 \acro{EDE}{Encrypt-Decrypt-Encrypt \acroextra{(mode of operation)}}
 \acro{EEPROM}{Electrically Erasable Programmable Read-Only Memory}
 \acro{EM}{electro-magnetic}
 \acro{FFT}{Fast Fourier Transform}
 \acro{FCC}{Federal Communications Commission}
 \acro{FIR}{Finite Impulse Response} 
 \acro{FIVR}{Fully Integrated Voltage Regulator}
 \acro{FNN}{Feedforward Neural Network}
 \acro{FPGA}{Field Programmable Gate Array}
 \acro{FSK}{Frequency Shift Keying}
 \acro{GMSK}{Gaussian Minimum Shift Keying}
 \acro{GPIO}{General Purpose I/O}
 \acro{GPU}{Graphics Processing Unit}
 \acro{HD}{Hamming Distance}
 \acro{HDL}{Hardware Description Language}
 \acro{HF}{High Frequency}
 \acro{HMAC}{Hash-based Message Authentication Code}
 \acro{HW}{Hamming Weight}
 \acro{IC}{Integrated Circuit}
 \acro{ID}{Identifier}
 \acro{ISM}{Industrial, Scientific, and Medical \acroextra{(frequencies)}}
 \acro{IIR}{Infinite Impulse Response}
 \acro{IP}{Intellectual Property}
 \acro{IoT}{Internet of Things}
 \acro{IV}{Initialization Vector}
 \acro{JTAG}{Joint Test Action Group}
 \acro{LF}{Low Frequency}
 \acro{LFSR}{Linear Feedback Shift Register}
 \acro{LLM}{Large Language Model}
 \acro{LQI}{Link Quality Indicator}
 \acro{LSB}{Least Significant Bit}
 \acro{LSByte}{Least Significant Byte}
 \acro{LUT}{Look-Up Table}
 \acro{MAC}{Message Authentication Code}
 \acro{MF}{Medium Frequency}
 \acro{MMU}{Memory Management Unit}
 \acro{MITM}{Man-In-The-Middle}
 \acro{MSB}{Most Significant Bit}
 \acro{MSByte}{Most Significant Byte}
 \acro{MSK}{Minimum Shift Keying}
 \acro{MSR}{Model Specific Register}
 \acro{NLFSR}{Non-Linear Feedback Shift Register}
 \acro{NLF}{Non-Linear Function}
 \acro{NN}{Neural Network}
 \acro{NFC}{Near Field Communication}
 \acro{NRZ}{Non-Return-to-Zero \acroextra{(encoding)}}
 \acro{NVM}{Non-Volatile Memory}
 \acro{OOK}{On-Off-Keying}
 \acro{OP}{Operational Amplifier}
 \acro{OTP}{One-Time Password}
 \acro{PC}{Personal Computer}
 \acro{PCB}{Printed Circuit Board}
 \acro{PhD}{Patiently hoping for a Degree}
 \acro{PKE}{Passive Keyless Entry}
 \acro{PKES}{Passive Keyless Entry and Start}
 \acro{PKI}{Public Key Infrastructure}
 \acro{PMBus}{Power Management Bus}
 \acro{PoC}{Proof-of-Concept}
 \acro{POR}{Power-On Reset}
 \acro{PPC}{Pulse Pause Coding}
 \acro{PRNG}{Pseudo-Random Number Generator}
 \acro{PSK}{Phase Shift Keying}
 \acro{PWM}{Pulse Width Modulation}
 \acro{RAPL}{Running Average Power Limit}
 \acro{RDP}{Read-out Protection}
 \acro{RF}{Radio Frequency}
 \acro{RFID}{Radio Frequency IDentification}
 \acro{RKE}{Remote Keyless Entry}
 \acro{RNG}{Random Number Generator}
 \acro{ROM}{Read Only Memory}
 \acro{ROP}{Return-Oriented Programming}
 \acro{RSA}{Rivest Shamir and Adleman}
 \acro{SEV}{Secure Encrypted Virtualization}
 \acro{SCA}{Side-Channel Analysis}
 \acro{SDR}{Software-Defined Radio}
 \acro{SGX}{Software Guard Extensions}
 \acro{SNR}{Signal to Noise Ratio}
 \acro{SHA}{Secure Hash Algorithm}
 \acro{SHA-1}{Secure Hash Algorithm 1}
 \acro{SHA-256}{Secure Hash Algorithm 2 (256-bit version)}
 \acro{SMA}{SubMiniature version A \acroextra{(connector)}}
 \acro{SMBus}{System Management Bus}
 \acro{I2C}{Inter-Integrated Circuit}
 \acro{SPA}{Simple Power Analysis}
 \acro{SPI}{Serial Peripheral Interface}
 \acro{SPOF}{Single Point of Failure}
 \acro{SoC}{System on Chip}
 \acro{SVID}{Serial Voltage Identification}
 \acro{SWD}{Serial Wire Debug}
 \acro{TCB}{Trusted Computing Base} 
 \acro{TDX}{Trust Domain Extensions}
 \acro{TLB}{Translation Lookaside Buffer}
 \acro{TEE}{Trusted Execution Environment}
 \acro{TMTO}{Time-Memory Tradeoff}
 \acro{TMDTO}{Time-Memory-Data Tradeoff}
 \acro{TZ}[TrustZone]{TrustZone}
 \acro{RSSI}{Received Signal Strength Indicator}
 \acro{SHF}{Superhigh Frequency}
 \acro{UART}{Universal Asynchronous Receiver Transmitter}
 \acro{UHF}{Ultra High Frequency}
 \acro{UID}{Unique Identifier} 
 \acro{USRP}{Universal Software Radio Peripheral}
 \acro{USRP2}{Universal Software Radio Peripheral (version 2)}
 \acro{USB}{Universal Serial Bus} 
 \acro{VHF}{Very High Frequency}
 \acro{VLF}{Very Low Frequency}
 \acro{VHDL}{VHSIC (Very High Speed Integrated Circuit) Hardware Description Language}
 \acro{VM}{Virtual Machine}
 \acro{VR}{Voltage Regulator}
 \acro{WLAN}{Wireless Local Area Network}
 \acro{XOR}{Exclusive OR}
 \acro{IR}{Intermediate Representation}
 \acro{OCD}{On-Chip Debug}
 \acro{OS}{Operating System}
 \acro{ML}{Machine Learning}
 \acro{DVFS}{Dynamic Voltage and Frequency Scaling}
 \acro{MMIO}{Memory-Mapped I/O}
\end{acronym}

\acused{AES}
\acused{CRC}
\acused{DES}
\acused{EEPROM}
\acused{RSA}
\acused{USB}
\acused{SHA-1}
\acused{IC}
\acused{CPU}
\acused{USB}
\acused{DRAM}
\acused{ARM}
\acused{AMD}

\title{Activation Functions Considered Harmful: \\
Recovering Neural Network Weights through Controlled Channels}

\author{
    \IEEEauthorblockN{Jesse Spielman
    \IEEEauthorblockA{\textit{School of Computer Science} \\
    \textit{University of Birmingham}\\
    Birmingham, UK \\
    jxs1366@bham.ac.uk}
    \and
    \IEEEauthorblockN{David Oswald}
    \IEEEauthorblockA{\textit{School of Computer Science} \\
    \textit{University of Birmingham}\\
    Birmingham, UK \\
    d.f.oswald@bham.ac.uk}
    \IEEEauthorblockA{\textit{School of Computer Science} \\
    \textit{Durham University}\\
    Durham, UK \\
    david.f.oswald@durham.ac.uk}
    \and
    \IEEEauthorblockN{Mark Ryan}
    \IEEEauthorblockA{\textit{School of Computer Science} \\
    \textit{University of Birmingham}\\
    Birmingham, UK \\
    m.d.ryan@bham.ac.uk}
    \and
    \IEEEauthorblockN{Jo Van Bulck}
    \IEEEauthorblockA{\textit{DistriNet} \\
    \textit{KU Leuven, Belgium}\\
    Leuven, Belgium \\
    jo.vanbulck@cs.kuleuven.be}
    }
}

\maketitle
\begin{abstract}

Recent advancements in hardware-based enclaved execution environments, such as Intel SGX, aim to protect critical model parameters in high-stakes machine learning applications increasingly moving to end-user or cloud environments.  However, this also introduces the risk of privileged side-channel attacks, traditionally aimed mainly at cryptographic targets.

In this paper, we develop a novel attack methodology that exploits input-dependent memory access patterns in common neural network activation functions to extract hidden model parameters.  In several case studies using the SGX-Step attack framework and the Tensorflow Microlite library, we demonstrate complete recovery of first-layer weights and biases, along with partial recovery of deeper layer parameters under specific conditions.  Our novel attack technique requires only 20 queries per input per weight to obtain all first-layer weights and biases, with an average absolute error of less than 1\%, improving over prior model stealing attacks.  Furthermore, a broader ecosystem analysis reveals the widespread use of activation functions with input-dependent memory access patterns in popular machine learning frameworks and maths libraries. Our findings highlight the limitations of deploying confidential models in SGX enclaves and emphasise the need for stricter side-channel validation of machine learning implementations, akin to the vetting efforts applied to secure cryptographic libraries.

\end{abstract}

\begin{IEEEkeywords}
Machine Learning Security, Side Channel Attacks, Trusted Execution, Confidential Computing, SGX
\end{IEEEkeywords}

\section{Introduction}
Training and deploying \ac{ML} models, in particular \acp{NN}, often requires a massive investment in computational resources. Hence, the threat of model stealing attacks is significant, because they allow an attacker to abscond with an \ac{ML} application, causing substantial financial damage as well as security and privacy risks.  Attacks could be used for industrial espionage (in the case of a network that solves a difficult problem, \eg a sophisticated automated trading system), or to ease the creation of adversarial examples to defeat a classification system, \eg for spam filtering or intrusion detection.  An application that incurs a fee to access per interaction could also be duplicated to remove this restriction.  While existing model stealing attacks show how to duplicate \acp{NN} essentially via brute force, they scale poorly for increasingly complex networks and require detailed access to output probabilities or physical access for measurements such as power consumption.

Because \ac{ML} (both training and inference) requires substantial computation, there is a strong focus on performance.  Further, because of the hardware requirements of \ac{ML} training and inference (such as in the case of an AI model which backs an end user-accessible web interface), there is an increasing need to host \ac{ML} workloads in the cloud.  However, this move onto (often shared) hardware introduces a number of security concerns.  To mitigate the risk, substantial work has been done using hardware-based \acp{TEE} such as Intel \ac{SGX} or AMD \ac{SEV} to protect confidential \ac{ML} workloads~\cite{russinovich2023confidential,intel_secure_ml,kumar23,li2024coreguard,kethireddy2024secure}.
These TEEs are presumed to facilitate the secure and efficient outsourcing of critical \ac{ML} computations to third-party hardware, while simultaneously preserving the confidentiality of model parameters that embody key \ac{IP}.

A core component of \acp{NN} are their activation functions, chosen by developers (alongside other architectural decisions such as the number of layers or the learning rate) to introduce non-linearity.  These functions are generally computed once for each neuron based on the sum of the multiplication of all incoming inputs by their weights (plus a bias term).  Often, these functions are provided by a \ac{ML} framework such as Tensorflow or PyTorch, which include routines for training, running inference, and deploying models.
It is important to note that these framework-provided activation functions usually rely on basic mathematical functions such as \code{std::max()} or \expfunc{}, which are often provided by the system's standard libraries (\eg \code{glibc}) and optimised for speed, not side-channel resistance.  \acp{TEE} like Intel \ac{SGX} often supply their own (restricted) standard library (\eg \ac{SGX}'s \code{tlibc}), which provides implementations known to work within the special enclaved environment.  Thus, \ac{ML} frameworks inherit some security properties from underlying standard library maths functions.

Research into \ac{SCA} (both with physical access as well as remotely exploitable side channels like timing) has led to the ``hardening'' of certain security sensitive algorithms, especially in cryptographic libraries. However, less attention has been paid to the side-channel security of \ac{ML} libraries, even though their parameters (weights and biases) are today often akin to cryptographic secrets in sensitivity. Especially in the context of \ac{TEE}-protected \ac{ML} workloads, the strengthened side-channel adversary model has not been sufficiently studied.

In this paper, we show how pervasive data-dependent memory access patterns in activation functions across many \ac{ML} frameworks can lead to the deterministic recovery of hidden model parameters when deployed in a \ac{TEE}.
Particularly, we develop a novel methodology to recover partial weights and biases from the first (and deeper) layer(s) of a victim network when provided with an instruction-granular page-access trace.

As a practical case-study, we deploy three \tfulite{} models inside \ac{SGX} enclaves and extract deterministic, instruction-granular page access traces using the SGX-Step~\cite{van_bulck_sgx-step_2017} framework.  Our first case study demonstrates how we can use those traces to successful recovery all first layer weights and biases to more than 5 decimal places of accuracy with 55 invocations of the network, or with less than 1\% average error with around 20 invocations after an initial calibration phase.  Our end-to-end attack outperforms \citeauthor{tramer_stealing_2016} model stealing attacks~\cite{tramer_stealing_2016}, which require around 100 invocations for each parameter in simpler classification networks.  We then use our second and third case studies to explore deeper layers and much larger networks, respectfully.  Considering the wider ecosystem, we survey widely used \ac{ML} libraries and find widespread use of secret-dependent access patterns in activation functions.  Furthermore, we discuss to what extent our attack vector applies to other TEEs.

\mypar{Contributions} Summarised, our contributions are:

\begin{itemize}
    \item A novel methodology to recover weights and biases from partial memory-access side-channel traces.

	\item Three end-to-end case studies demonstrating concretely how, by exploiting our memory-trace extraction technique, we can perform accurate weight/bias recovery from \tfulite{} networks running inside SGX enclaves in a few different configurations.

    \item A comprehensive survey of input-dependent accesses in common activation functions across popular ML and standard libraries.

	\item An open-source \tfulite{} \ac{SGX} benchmark for future work on attacks and mitigations.
\end{itemize}

\mypar{Ethics and Open Science}
All experiments were conducted on \ac{PoC} implementations on our own local machines.
To ensure the reproducibility of our results, and to enable future research on side-channel attacks and defenses, we release all code and data as open-source at \url{https://github.com/heavyimage/afch_paper}.

\mypar{Scope}
To the best of our knowledge, this is the first study to thoroughly analyse side-channel implications of deploying ML workloads in TEEs, which are increasingly being suggested by industry and academia to preserve \ac{IP} and end-user privacy.
In this paper, we do not claim a fully-fledged, optimised and weaponised real-world attack; instead we focus on developing a \emph{novel attack methodology} to steal partial model parameters via case-study-driven evaluation.
While full recovery of real-world networks may be out of reach, we show that motivated adversaries can fully automatically extract thousands of hidden, supposedly secure parameters.
Following the rich tradition in cryptography, where a long line of increasingly potent side-channel attacks have highlighted the need for constant-time programming practices, we hope our work will draw attention to the security and privacy trade-offs of using TEEs for private \ac{ML} and underline the importance of side-channel-resistant coding practices.

\section{Background and Related Work}
\label{sec:background}

\mypar{\aclp{TEE}}
One of most widely studied \acp{TEE} is Intel's \ac{SGX}, which, even though discontinued for client CPUs, is widely supported on Xeon Scalable server \acp{CPU} and now marketed for use cases such as secure \ac{ML} deployments~\cite{intel_secure_ml}.  \ac{SGX}-secured code is executed in a hardware-protected \emph{enclave} (with its memory encrypted) such that even the \texttt{root} user cannot gain access.
\ac{SGX} also allows secrets to be securely loaded into the enclave remotely, and subsequently be (un)sealed for local storage. This allows data of the enclave (such as a pre-trained \ac{ML} model) to be also protected while at rest.
SGX does not provide guarantees against timing or memory-based side channels~\cite{sgxdeveloperguide}, rendering enclaves vulnerable to cache and page fault attacks~\cite{nilsson_survey_2020,xu_controlled-channel_2015}. Moreover, due to SGX’s privileged adversary model, such attacks can achieve high temporal resolution using tools like \sgxstep{}~\cite{van_bulck_sgx-step_2017}, which exploit timer interrupts to single-step enclave execution, effectively interleaving enclave execution with attacker-controlled code at an instruction-level granularity.

Other \acp{TEE} of note include \acs{ARM}'s \acl{TZ} which brings the concept of \acs{TEE} to mobile devices~\cite{arm_tz}.  Intel's \ac{TDX}~\cite{intel_tdx} and \acs{AMD}'s \ac{SEV}~\cite{amd_sev} are \acp{TEE} that provide hardware-backed isolation guarantees at the level of entire \acp{VM}.  

\mypar{Model Stealing Attacks}
The recovery of parameters from a trained neural network via ``model stealing'' is well documented in the literature.  Such practical recovery attacks, first demonstrated by \citeauthor{tramer_stealing_2016}, enabled the duplication (stealing) of neural networks and other models~\cite{tramer_stealing_2016}.   In these attacks, the original network is used as an oracle to train a duplicate network with similar hyperparameters.  The attack works by choosing inputs via a search of the parameter space of the network (using one of several different heuristics) and joining that input with the corresponding output probabilities which are returned.  These data pairs are used as training examples for the new network.  ``Solving'' a network in this way results in the calculation of hidden parameters that are derived from but not necessarily precise to the original source network.  Furthermore, they require a large number of queries, around 100 per parameter, as well as access to the output probabilities to essentially brute-force the space of the network.  While these attacks may reconstruct the overall structure and general features of the network, it remains an open question how effective they are at reproducing the specifics of the original network, \eg when attempting to transfer adversarial examples from duplicate to original.  An alternative approach by \citeauthor{canales2024polynomial} (and references therein) treat the recovery of parameters as cryptanalysis, showing that the \emph{sign} information for each neuron in a 1.2M parameter CIFAR10 network can be recovered in around 30\,min using a 40\,GB A100 \ac{GPU}~\cite{canales2024polynomial}. 

\mypar{Side-Channel Attacks on \ac{ML} Implementations}
An alternative approach to recover model parameters is to use a software-based (timing or cache) side channel. Most research in this direction has focused on the recovery of hyperparameters, such as the network's architecture, but \emph{not} the precise weights: \citeauthor{yan_cache_2020} demonstrate a technique for recovering the hyperparameters of a victim network such as the number of layers and their activation functions using a Flush+Reload cache attack~\cite{yan_cache_2020}. Other works in this direction have used similar cache attacks~\cite{hong2020security}, execution time~\cite{duddu2019stealing}, looked at embedded devices~\cite{Won21}, and considered equivalent approaches on \acp{GPU}~\cite{Naghibijouybari18,Wei20,Naghibijouybari21}. Other uses for software-based \ac{SCA} include the recovery of network inputs from floating point timing~\cite{Dong19}, membership inference~\cite{ali2022unintended}, and the generation of adversarial examples~\cite{Nakai_Suzuki_Fujino_2021,YuichiroDan2021}.

Because the targeted leakages are much smaller and hence harder to exploit, less attention has been paid to full model recovery, \ie weights and biases, through software-based side channels:
\citeauthor{Alder20} show how an attacker-controlled configuration of the x86 floating point unit can be used to recover weights from a toy \ac{NN} (using a custom implementation) running in an \ac{SGX} enclave~\cite{Alder20}.

\citeauthor{Gongye20} show that, if an adversary obtains the precise timing of each network layer, in certain cases they can use timing leakages from floating point to recover the full model~\cite{Gongye20}. In particular, the authors rely on different execution timing on certain x86 CPUs when processing ``subnormal'' numbers. \citeauthor{Gongye20} do not specify whether their attack applies to real-world \ac{ML} libraries, and assume that the necessary precise timing measurements ``can be achieved by either analysing the cache access pattern [...] or through visual inspection of power traces''. In contrast to their work, in this paper, we consider the input-dependent memory access patterns of activation functions in widely used \ac{ML} libraries, showing an end-to-end attack that uses \sgxstep{} for reliable side-channel observations.

In contrast, side-channel attacks that recover the full model with physical access to the hardware and measurements of \ac{EM} emanation or power consumption have been further developed~\cite{Hua18,Batina19,Yu20,horvath2024barracuda}. However, as these attacks require measurements on or in close proximity to the targeted chips, they are of less relevance to datacenter applications, where physical access to the underlying servers is typically closely guarded.
Hence, in this paper, we focus on software-based side-channel attacks to recover weights/biases without physical access and without access to the output probabilities.

\mypar{Controlled Channel Attacks} For \ac{SGX}, Intel explicitly places the burden of preventing secret-dependent memory accesses for security-critical code on the developer~\cite{Wenhao17}.
In practice, this means that enclave developers should avoid data-dependent patterns of the form \code{array[x]} or \code{if(x) func()}, where \code{x} depends on a secret.
It is important to note that in practice, such secret-dependent access patterns may only become evident at the machine code level, as optimising compilers might \eg opt to compile ``branchless'' code into code with conditional branches, and vice versa.

\begin{figure}[tbp]
    \includegraphics[width=\linewidth]{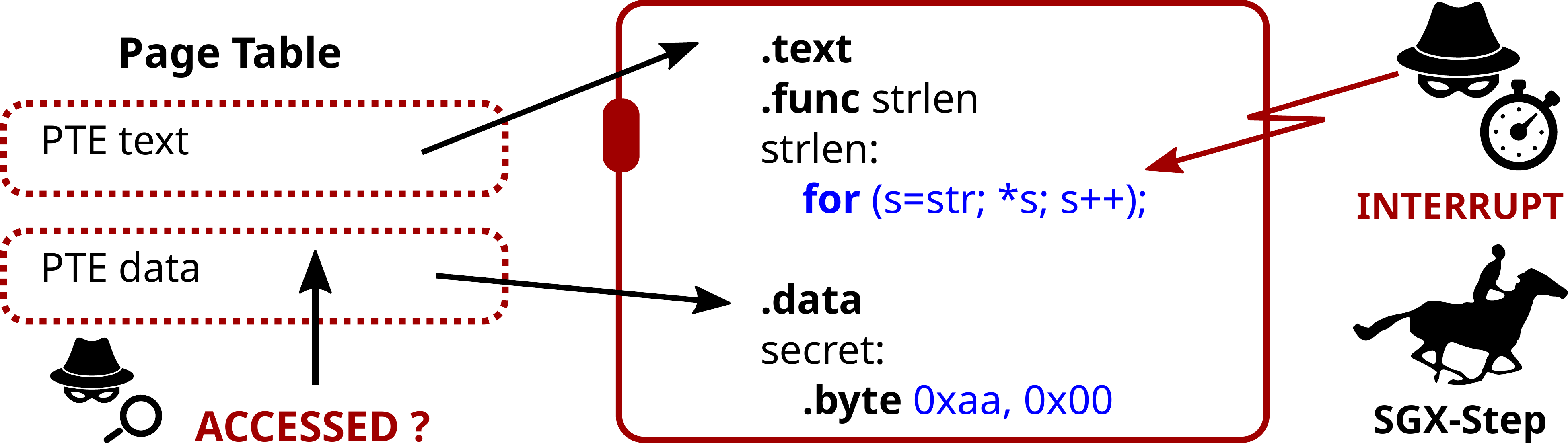}
	\caption{Victim enclaves exhibiting tight memory-access patterns can be precisely interrupted at instruction-level granularity using \sgxstep{} allowing to extract deterministic page-access count traces.
    }
    \label{fig:cc_sgxstep}
\end{figure}

Apart from well-known microarchitectural cache leakage~\cite{Brasser17}, which is notoriously noisy, a number of \ac{SGX}-specific side channels were discovered that exploit the privileged attacker's control over the untrusted \ac{OS}.
One of the first ones being ``controlled channels'' as introduced by \citeauthor{xu_controlled-channel_2015}~\cite{xu_controlled-channel_2015}. This method exploits the fact that page faults within an \ac{SGX} enclave are handled by the untrusted \ac{OS}. A privileged adversary can, hence, temporarily unmap a page, observe whether a page fault occurs, and from this fully \emph{deterministically} infer enclave memory access patterns at a 4\,KiB, page-level spatial granularity.  While such page-fault attacks have been proven particularly powerful, \eg to extract text and images~\cite{xu_controlled-channel_2015} or full cryptographic keys~\cite{shinde2016preventing}, their relatively coarse-grained spatial resolution may limit exploitability.  Consider the example code snippet provided in \Cref{fig:cc_sgxstep}, where a tight \texttt{strlen} loop is executed that fits entirely within a single code and data page.  As subsequent accesses to the same page are cached in the processor's \ac{TLB}, and the enclave needs both the code and data page to make forward progress, page-fault adversaries only observe the first access to a page and are not able to distinguish successive \texttt{strlen} loop iterations.  To overcome this limitation, \citeauthor{VanBulck17} showed that privileged adversaries may also monitor page-table attributes~\cite{VanBulck17} using the `accessed' and `dirty' bits to count accesses to a certain page.

The open-source \sgxstep{} framework~\cite{van_bulck_sgx-step_2017,vanbulck2023sgxstep} allows for precise single-stepping of production enclaves using privileged x86 APIC timer interrupts, such that page access patterns can be \emph{deterministically}  monitored for every enclave instruction, giving precise insights into the operation of a victim enclave.  By conservatively {under-estimating} the APIC timer interval, \sgxstep{} results in either zero or single-steps, but \emph{always} avoids multi-steps \cite[Table~5]{vanbulck2023sgxstep}. Subsequent works \cite{VanBulck18, moghimi2020copycat, aldaya2020one, van_bulck_tale_2019, constable2023aexnotify} showed that the accessed (A) bit in the enclave code page \textit{deterministically} distinguishes single-steps (enclave instruction retired; A=1) from zero-steps (A=0). 
This ability to perfectly and deterministically single-step production enclaves using \sgxstep{} prompted Intel to develop the opt-in AEX-Notify hardware-software mitigation (cf.\ \cref{sec:discussion}).
To date, instruction-granular page-access traces extracted with \sgxstep{} have been repeatedly abused to reliably exploit enclave interface vulnerabilities~\cite{van_bulck_tale_2019} or to reconstruct cryptographic key material~\cite{moghimi2020copycat,aldaya2020one}.
In the following, we show that such traces are sufficient to recover weights from real-world \ac{NN} libraries running inside a production \ac{SGX} enclave. 

\section{System and Adversary Model} \label{sec:preliminaries}

\mypar{Attacker Capabilities}
We adhere to Intel's standard \ac{SGX} threat model, where the adversary has full \code{root} access on the victim machine, as is \eg the case when deploying an \ac{ML} model on external cloud infrastructure.
The attacker is a software-only adversary without local physical access, for instance they may gain root access remotely over the network.
Such privileged software attackers can send arbitrary inputs through the enclave software interface and can further interrupt the enclave by manipulating page tables and interrupts (cf.\ \cref{sec:background}).
For the latter, we leverage the widely-used \sgxstep{}~\cite{van_bulck_sgx-step_2017} framework to precisely single-step a target enclave and obtain instruction-granular page access traces.
Notably, while attackers may debug enclaves using a copy of the victim code to facilitate attack development, we conduct all final attacks with \sgxstep{} on target enclaves running in SGX production mode, i.e., without access to debug features.

While our work focuses on the Intel SGX architecture, the underlying insights and weight-recovery techniques are applicable to other TEEs.  Notably, the required side-channel primitives offered by SGX-Step, \ie precise single-stepping and page accesses, have been demonstrated on alternative VM-based TEEs like AMD SEV~\cite{Wilke_Wichelmann_Rabich_Eisenbarth_2023} and Intel TDX~\cite{Wilke24, shome2024tdexit-notify}.

\mypar{Target Model} We assume a pre-trained \ac{ML} network.  We do not consider the training process, which is orthogonal to our attack (but we believe it to be an interesting avenue for future research).  We focus on \acp{FNN} because they effectively illustrate the problem we exploit and are easily deployable in \ac{SGX} enclaves (\eg without dynamic linking or requiring \code{libc}) using \tfulite{}.  Lastly, we assume the network architecture (at least the number of neurons per layer and activation functions) is known, \eg it is an open-source or well-known architecture or a transfer-learning model, or can be recovered via a cache attack~\cite{yan_cache_2020}.

\mypar{Target Enclave} We assume that the network has been securely loaded into a production-mode \ac{SGX} enclave to perform ``secure'' inference.  The network's weights and biases are confidential (sealed), while the enclave source code is known.  The latter is a common assumption in SGX attacks~\cite{xu_controlled-channel_2015,van_bulck_sgx-step_2017,VanBulck17,Brasser17,shinde2016preventing} and the default case in the Intel SGX architecture and SDK.  Non-standard confidential-code deployments only add ``security through obscurity'' by requiring an additional, out-of-scope reversing phase~\cite{puddu2024lack,yan_cache_2020}.  Since the attacker never has access to a self-contained binary that contains the hidden parameters, an attack such as the one proposed by \citeauthor{liu2023decompiling} is not possible \cite{liu2023decompiling}.  Inputs are passed into the enclave and the output is returned outside the enclave through the standard \texttt{ecall} interface.  Based on the security guarantees of \ac{SGX}, an \ac{ML} developer would assume that their network is protected against duplication, as inference is handled entirely within an enclave and weights are never stored outside.

\begin{figure*}[htbp]
    \vspace*{-2mm}
    \includegraphics[width=\textwidth]{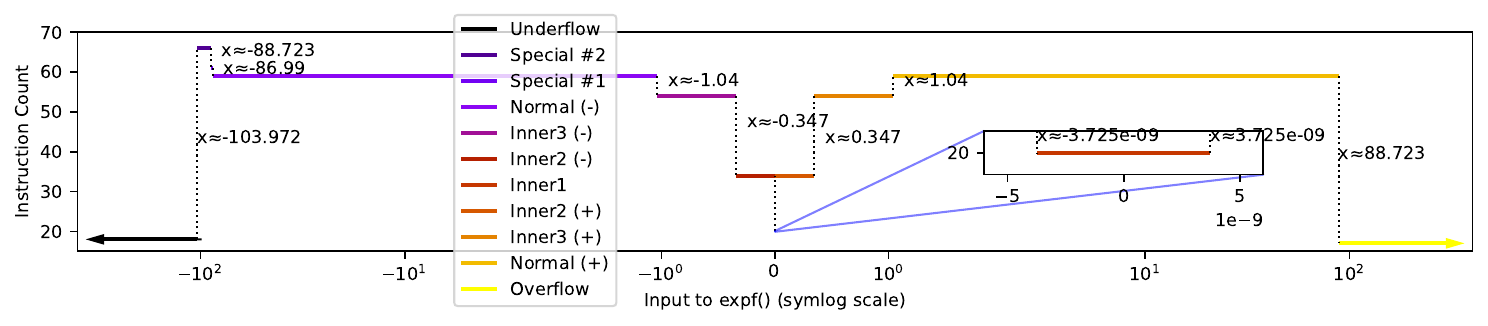}
	\vspace*{-7mm}
	\caption{Visualization of \expffunc{} CPU instruction count regions by input.  The Underflow and Overflow regions have different instruction counts: 18 and 17 respectively.  Note also the symmetrical regions of Inner1, inner2, and Inner3.}
    \label{fig:expf_state_map}
	\vspace*{-5mm}
\end{figure*}

\label{sec:vulnerable_functions}
\mypar{Activation Functions} Finally, our attack requires that the victim network use a vulnerable activation function which contains at least two \emph{input-dependent branches with unique step counts} such as the time they take to return. To give an example, \Cref{lst:expf_in_tlibc} shows some different code paths of the implementation of \expffunc{} in \ac{SGX}'s \code{tlibc} \cite{sun_microsystems_tlibcs_1993}.  Observing the memory access/execution patterns, \eg with \sgxstep{}, allows an attacker to learn which return case an input results in, and thus obtain information on the input to \expfunc{}.

For this reason, we mostly consider activation functions that include a call or calls to a functions in the \expfunc{} family, \eg \sigmoidfunc{}, \tanhfunc{}, and \softmaxfunc{}.  Notably, such activation functions are widely used in transformer architectures (even though we do not consider attacks on transformers in this paper), as with \sigmoidfunc{} in Deepseek V3 \cite{liu2024deepseek}, \gelufunc{} in BERT \cite{devlin2019bert} and \swiglufunc{} \cite{shazeer2020gluvariantsimprovetransformer} in Llama \cite{touvron2023llama}.  \softmaxfunc{} was specified in the initial transformer paper \cite{vaswani2017attention} and is still used in many of these implementations.
Note that while several variants of \expfunc{} implementations exist, they commonly feature the same structure, where multiple error cases are handled with early return statements.

\begin{lstlisting}[label={lst:expf_in_tlibc}, caption={Excerpt from \expffunc{} in \code{tlibc}.}, captionpos=b, escapechar=§]
  if(hx >= 0x42b17218) {			/* if |x|>=88.721... */
  	if(hx>0x7f800000) return x+x; /* NaN */
  	if(hx==0x7f800000) return (xsb==0)? x:0.0;
  	if(x > o_threshold) return ...; // overflow
  	if(x < u_threshold) return ...; // underflow
  } if(hx > 0x3eb17218) {		
  	if(hx < 0x3F851592) { /* ... */ } else { /* ... */ }
  } else if(hx < 0x31800000)  {	/* when |x|<2**-28 */
  	if(huge+x>one) return ...; // inexact 
  }
\end{lstlisting}

Attacking an activation function containing a call to \expfunc{} is similar to attacking \expfunc{} directly.  One must be careful to note the sign of the input (as in the case of the $-x$ in \sigmoidfunc{}'s $\exp\left(-x\right)$) and search in the correct direction.  In the case of functions like \code{tanh()}, which can include multiple discrete calls to \expfunc{}, the code to process the memory access traces cannot assume a one-to-one relationship between a call to \expfunc{} and a single neuron's activation.  Another important consideration is the expressive power of the function, which we discuss in more detail in \Cref{sec:deeper-layers}.

We note \acp{NN} may access the maths functions that underpin their activations from a few different sources.  Whether defined by an \ac{ML} framework itself or imported externally from low level standard libraries such as \code{glibc}'s \code{libm} or even dedicated high performance maths libraries like \code{Eigen} (which may be vectorised to exploit hardware acceleration features), the code may be vulnerable to our attack.  We explore this with our second case study in \Cref{sec:pocII} and will more generally consider how the source of these functions can affect their vulnerability to our attack in \Cref{sec:ecosystem_analysis}.

Finally, despite our focus on the \expfunc{} family of functions, any function exhibiting measurable input-dependent variation in instruction counts (or memory access patterns) is likely vulnerable.  As long as the attacker can force a vulnerable function to process inputs that fall into multiple distinguishable classes, it is possible to search for the boundary points between these cases, as we discuss in detail in \Cref{sec:methodology}.  In particular, \relufunc{} does not rely on a call to \expfunc{}, but on conditional logic or a call to \code{std::max()}.  We summarise common activation functions in \Cref{tab:function_vuln} and specifically discuss the vulnerability of \relufunc{} in greater detail in \Cref{sec:relu}.

\section{Methodology} \label{sec:methodology}
First, we describe the overall concept for our attack (\Cref{sec:attack-concept}) against a \ac{NN} and then consider (\Cref{sec:framework_functions}) different ways \ac{ML} frameworks may implement activation functions.  Next (\Cref{sec:collecting_traces}), we show how to perform our attack practically using \sgxstep{} to generate per-neuron traces.  We then explain in more detail how to attack the first layer (\Cref{sec:first_layer}) and the subsequent layers (\Cref{sec:deeper-layers}).  Finally (\Cref{sec:estimating_attack_performance}) we estimate the performance of our attack.

\subsection{Attack Concept}\label{sec:attack-concept}
The starting point for our attack is the above observation (\cf \Cref{sec:vulnerable_functions}) that, unlike in cryptographic libraries, the maths functions that underpin many \ac{ML} activation functions have not been hardened to ensure they execute in constant time and without input-dependent branches and memory access patterns.  A classic example in the cache timing literature is a \acp{LUT} that may be accessed in deterministic patterns based on the function inputs.  We believe this work has not been undertaken in the context of \ac{ML} due to performance concerns or because security is seen as less of a focus.

By examining the underlying functions statically, dynamically, or by reading the code (if available), we can deduce the possible branches through the function and the timing as well as memory access patterns therein.  Also of interest are cases where certain input values are handled differently (via multiple \code{return} statements).  A special case of those are \emph{early returns}, where certain inputs cause the function to return before the full output is computed.

\begin{table}[htbp]
	\caption{Survey of Potentially Vulnerable Activation Functions.} 
	\label{tab:function_vuln}
	\centering
	\small
	\setlength\tabcolsep{2pt}
	\resizebox{\columnwidth}{!}{
		\begin{tabular}{llll}
			\toprule
			\bf Activation Function     & \bf Output Range                  & \bf Contains \expfunc{}     & \bf Contains \code{max()}\\
			\midrule                                                        
			\sigmoidfunc{}              & $(0,1)$                           & \cmark                      & \xmark \\
			\tanhfunc{}                 & $(-1,1)$                          & \cmark                      & \xmark \\
			\softplusfunc{}             & $(0, \infty)$                     & \cmark                      & \xmark \\
            \code{ELU()}                & $( -\alpha, \infty)$              & \cmark                      & \xmark \\
            \code{SELU()}               & $( -\lambda * \alpha, \infty)$    & \cmark                      & \xmark \\
			\midrule                                                        
			\relufunc{}                 & $(0,\infty)$                      & \xmark                      & \cmark \\
			\bottomrule
		\end{tabular}
	}
	\\ \smallskip
	\scriptsize
\end{table}

Given that such code patterns take different numbers of instructions (and exhibit different memory access patterns) depending on the input, more generally there is a correlation between different code paths and the execution trace.  Though these differences might be subtle (\eg a single instruction), they are measurable given a suitable side channel. Thus, if an attacker can determine which branch a function has taken, they learn something about the input or argument to that function.

As stated, one function with these properties is \expffunc{}.  \Cref{fig:expf_state_map} shows the 11 different return cases of this function based on input and their associated CPU instruction counts (collected using \code{gdb}).  We focus here on the \expffunc{} implementation from \acs{SGX}'s \code{tlibc}, but we note that we found other optimised implementations all using a similar algorithm and thus exhibiting similar input-dependent memory access patterns.  For ease of discussion we have labeled these 11 cases using the names shown in \Cref{fig:expf_state_map}.  For example, \code{expf(4)} results in case Normal $(+)$, whereas \code{expf(110)} results in Overflow.  We define a \emph{threshold} (shown in \Cref{fig:expf_state_map} as dotted black lines) as an input value that is on the precise border of two timing classes.  For example, the value \expfmaxbound{} is between Normal $(+)$ and Overflow.

We can also use knowledge of these regions and thresholds in the other direction: for example, if we know that an execution of \expffunc{} takes 17 instructions (in the Overflow region), we know the input to that function call must be greater than \expfmaxbound.  We note that the cases for Overflow and Underflow are both early returns and as such their instruction counts are much lower than the other `normal' returns which actually perform some sort of unique computation.

We can use these threshold values to leak information about the precise value of a partially controlled input.  If we consider the scenario where an input to \expffunc{} is being multiplied by a hidden constant $c$ and we have access to the instruction count of the execution, we can reveal $c$ by finding (via \eg a binary search) a threshold value between two instruction count classes and then dividing that value by our input.  For example, if we find that the input $3.2164$ multiplied by $c$ leads to threshold value \expfmaxbound, $c$ must be $\nicefrac{\expfmaxboundnb}{3.2164} \approx 27.5846 $.  We therefore define $3.2164$ as a \emph{convergence point}: an input to the system that causes a threshold value and thus a leak of the precise input to the function.

\begin{figure}[htbp]
	\small
	\begin{tikzpicture}[
            yscale=.5,
		init/.style={
			 draw,
			 circle,
			 inner sep=2pt,
			 font=\Huge,
			 join = by -latex
		},
		squa/.style={
			font=\Large,
			join = by -latex
		}
	]
	\begin{scope}[start chain=1]
		\node[on chain=1] at (0,1cm)  (x1) {$i_1$};
		\node[on chain=1,join=by o-latex] (w1) {$w_1$};
	\end{scope}
	\begin{scope}[start chain=2]
		\node[on chain=2] at (0,-1cm)  (x2) {$i_2$};
		\node[on chain=2,join=by o-latex] {$w_2$};
		\node[on chain=2,init] at (1.5cm,0)  (sigma) {$\displaystyle\Sigma$};
		\node[on chain=2,squa,label=above:{\parbox{2cm}{\centering Activation\\ function}}]   {$f(\Sigma)$};
		\node[on chain=2,squa,join=by -latex] {\small $output$};
	\end{scope}
	\node[label={[yshift=-0.2cm]above:Bias $b$}] at (sigma|-w1) (b) [yshift=.5cm]{};
	\draw[-latex] (w1) -- (sigma);
	\draw[o-latex] (b) -- (sigma);
	\end{tikzpicture}
	\caption{Visualization of a single 2-input neuron.  We use $\Sigma$ in this paper to represent the sum of all $i_i * w_i$ and $b$ passed into an activation function.}
	\label{fig:single_neuron}
\end{figure}

We now extend the above into an attack against a toy \ac{ML} network: consider a simple \ac{NN} with only a single neuron, as shown in \Cref{fig:single_neuron}.  The activation function is simply \expffunc{}.  The goal is to recover $w_1$, $w_2$, and $b$. We are free to choose values for $i_1$ and $i_2$.  We first note that the input to a neuron's activation function is the sum of the products of various user-controlled inputs and their hidden weights, plus a hidden bias term ($\Sigma$ in \Cref{fig:single_neuron}); this is similar to the case described previously except we have two hidden constants multiplied by two different inputs and an additional term, the bias, that is never multiplied with our inputs.

If there was no bias term, we could simply use a variant of the approach described above and iterate through each input (setting all other inputs to zero, allowing us to focus on one weight at a time) and recovering each parameter independently.  The addition of the bias term adds another unknown which means we can only find through constraints on the value of the hidden weights and bias:

\begin{equation*} \begin{aligned}
	i_1 * w_1 + 0 * w_2 + b = \expfmaxboundnb  
	\Rightarrow w_1 = \frac{\expfmaxboundnb - b}{i_1}
\end{aligned} \end{equation*}

To solve for all three unknowns, we find enough sets of values (one for each input to the neuron) to generate a system of equations to solve for as many unknowns as we have (number of weights plus bias term), for example:

\begin{equation*} \begin{aligned}
	i_1 * w_1 + \cancel{0 * w_2} + b &= threshold_1 \\
	\cancel{0 * w_1} + i_2 * w_2 + b &= threshold_2 \\
	i_3 * w_1 + i_4 * w_2 + b &= threshold_3 \\
\end{aligned} \end{equation*}

Here, we define $(i_1,\,0)$, $(0,\,i_2)$, and $(i_3,\,i_4)$ as \emph{convergence point sets}, because each set of chosen inputs forces the activation function to a threshold value.  Once we have enough equations (one for each unknown), we solve using linear algebra for the unknown weights and biases.  Note that the thresholds do not have to be different, though the more thresholds are hit, the more robust the solution.

Though it is possible to find convergence point sets like $(i_1,\,0)$ and $(0,\,i_2)$ above where all but one input is locked to zero and the remaining input is used to search for a threshold point, this is not required.  In fact, by locking all but one input to some small magnitude random numbers (\eg $i_3$ in the third equation) and searching on the single unlocked parameter ($i_4$), an arbitrary number of convergence point sets can be found.
Finally, if we replace \expfunc{} above with a different function (such as \sigmoidfunc{}) that also leaks threshold values, the only change is how to conduct the binary search; the recovery procedure is otherwise identical.  Building on this concept, we will show how it is possible to recover the hidden parameters of more complex networks to a high degree of accuracy.

\subsection{\ac{ML} Framework Functions} \label{sec:framework_functions}
Thus far we have largely assumed a low level mathematical definition for these functions.  However, as discussed in \Cref{sec:preliminaries}, a Tensorflow neuron with a \tanhfunc{} activation might not directly return the output of a call to the underlying maths library's implementation of \tanhfunc{}; there might be wrapping or processing of the result at the framework level.  This is of interest as we can also exploit input-dependent behaviour at this higher framework or application level.  An example of this is \tflite{}'s \sigmoidfunc{} function (called \code{Logistic()} in the source), shown in \Cref{sec:sigmoid_in_tflite}.  We can see that on top of any side-channel leaks due to calls to \code{std:exp()}, a higher-level set of memory access patterns and instruction count cases are introduced by a conditional routing to different functions (or simply a return of 0) based on the input.  In this way, this \sigmoidfunc{} function can actually be attacked at two levels which we will describe in \Cref{sec:pocII}.

\subsection{Capturing Traces with \sgxstep{}}\label{sec:collecting_traces}

To practically execute our attack, we used \sgxstep{} to collect the CPU instruction counts necessary to understand the state of our victim networks and apply the steps described above.  This required that our networks run inside an \ac{SGX} enclave.  Given the constraints of \ac{SGX}'s standard library replacement \code{tlibc}, we decided to employ \tfulite{}, a software package for running (converted) \tflite{} models on microcontrollers, instead of ``full'' Tensorflow (which relies on a variety of system calls not easily provided by \ac{SGX}).  Our \ac{SGX} enclave thus included a statically compiled \tfulite{} library to perform inference.  We converted our tensorflow models to \tflite{} and then further exported them as to a \tfulite{} byte array which can be compiled directly into a C program.  These byte arrays are added to the enclave code which allows us to perform ``secure'' inference inside \ac{SGX}.  We note that for practical deployments of \ac{SGX}-protected \ac{ML}, using such a reduced variant of Tensorflow would ease deployment, reducing code and thus \ac{TCB} size, while maintaining compatibility with Tensorflow-trained models.  The underlying implementation \cite{sun_microsystems_tlibcs_1993} of \expfunc{} used by our network during inference comes from SGX's \code{libc} alternative, \code{tlibc}.

We first profiled the execution of \tfulite{} on an attacker-controlled debug enclave, so that we could understand execution flow by directly observing call traces via the interactive \code{sgx-gdb} debugger.
An example can be seen in \Cref{sec:sgx_expf_stack}.  This provides an intuition for how \tfulite{} performs inference and which functions and pages are accessed.

\begin{figure*}[tbp]
    \includegraphics[width=\linewidth]{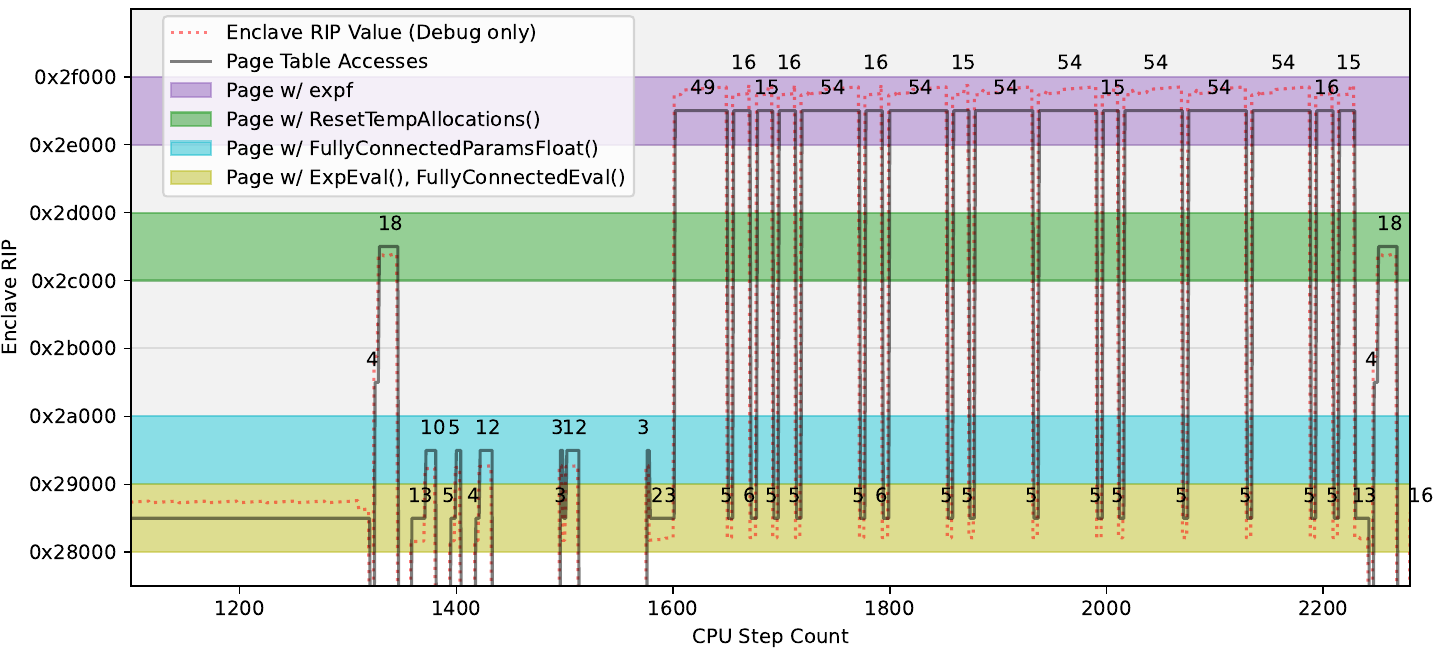}
	\caption{Inset of an \sgxstep{} instruction-granular page access trace showing the execution of the first layer of a \ac{NN} with 16 neurons.  Note that the intra-page instruction pointer (RIP) values (recovered by putting an enclave in debug mode) are displayed here only for reference purposes and are not used in our attack.  We include them to show that knowing the current page provides enough information to interpret traces without precise RIP values.}
    \label{fig:nn_runtime_exp_inset}
\end{figure*}

We then developed an attack application, which starts and interacts with the victim enclave in production mode, i.e., without access to debug features.
We use this to record page accesses which we later interpret into activation function states.  Using \sgxstep{}, we log all page accesses within the production enclave (after a certain trigger page was accessed) and interpret the results as part of a post-processing step.
Unlike \code{sgx-gdb}, \sgxstep{} only offers a relatively coarse-grained spatial resolution of 4\,KiB memory pages, which prevents  production enclave attackers from directly identifying the specific symbol (\eg function or data) accessed by the enclave code.
However, \sgxstep{}'s precise instruction-level temporal resolution enables the annotation of coarse-grained page-access traces with the exact number of instructions executed on each page.
As illustrated in \Cref{fig:nn_runtime_exp_inset}, this capability allows to accurately reconstruct the inference execution structure, even without knowledge of the lower 12-bit page offset of the secret enclave instruction pointer (ERIP).
By comparing the dotted red line which depicts the true ERIP values available to \code{sgx-gdb} and the black line which shows the reduced page-level granularity, one can see how little is lost.
It is worth nothing that each page can contain multiple functions and so it is difficult to assign a precise ``meaning'' to each page without context.  The instructions executed on one page could pertain to two very different activities at different points in execution.  However, the combination of coarse-grained page-access patterns along with the exact amount of instructions executed per page, annotated in black text, clearly suffices to identify individual function calls or data accesses.  In practice, this means that even if multiple symbols are co-located on the same 4\,KiB page, we can still interpret the logs.  We also know precisely the length (or length range) of different runs at different points during the execution, \ie we can also tell if a trace is invalid and immediately reject it.

By studying inference traces from the perspective of both \code{sgx-gdb} and \sgxstep{} (raw traces or visualizations like those shown in \Cref{fig:nn_runtime_exp_inset}), clear patterns emerge.  For example, the function \code{ExpEval()} is located on a different 4\,KiB page than \expffunc{} (each \code{0x1000} bytes is a new page) and we can plainly interpret the movement between them as the higher level neuron code invoking the lower level \expffunc{} maths function 16 times (once for each neuron in the first layer of the model this plot was constructed from).  The 18 instruction count long runs at the extreme left and right on page \code{0x2c000} seem to correspond to the start and end of a layer, which makes parsing each layer trivial.  We note that the step counts provided by \code{sgx-gdb} and \sgxstep{} may vary slightly (but deterministically)  due to the documented effects of macro-op fusion~\cite{moghimi2020copycat}.

The final part of our tooling is a Python script and library which further abstracts and manages invocations of the enclave program.  Because each neuron is accessed in order and each return occurs in sequence, it should be possible to monitor each neuron's step count independently.  In order to orchestrate the full attack, we use this script to interpret the log output by \sgxstep{} into a per-neuron state.  Even though we receive the state for every neuron in each log, we only focus on one at a time.  It takes about ten seconds to run each input through our \sgxstep{} setup and parse the trace.  An example output of the running attack is shown in \Cref{lst:neuron_centric_attack} in \Cref{sec:neuron_centric_attack}.

\subsection{Neuron-Centric Attack on First Layer}\label{sec:first_layer}

Here we discuss how to convert the exploit discussed in \Cref{sec:attack-concept} into a full attack on the first layer of a production \ac{NN}.  As before, the goal of this phase is to find a number of convergence point sets equal to the number of inputs plus one in order to have a system of equations we can solve to recover the weights and bias.

The first layer may be a special case if (without loss of generality) there is no filtering or normalization which pre-processes the input values.  If this is the case, we can choose any values we want to feed into the network (we cannot immediately inject values deeper into the network).  If there is filtering or normalization, however, we treat the first layer like all subsequent layers.

If we can inject arbitrary values into the first layer, we can conduct a chosen-input, neuron-centric attack pass by iterating over every neuron in the first layer and recovering their hidden parameters in 2 phases: calibration and binary search.

\subsubsection{The Calibration Phase}\label{sec:calibration}

Because the attack is a binary search to find hidden parameters of unknown value, can achieve dramatic speedups if we can find a reasonable upper bound to start the search from, rather than stepping down from the maximum possible float value.  It is also useful to know the signs of the weights connected to our neurons.  We can recover both pieces of information through the calibration step.

We prepare a test input to the network that is all small, non-zero values except for the first value which is a very large value.  Next, we perform inference on this test input and consult the trace to see if the target neuron overflowed or underflowed.  If it has, we know both that this input is already large enough to cause an overflow/underflow and the sign of the weight between the first input and this neuron.  If the neuron has not overflowed or underflowed, (owing to a comparatively lower magnitude weight) we simply multiply the large value in our test input by 10 and try again.  We continue in this way until we find the sign and magnitude required to over/underflow this neuron from each input to the network.  Note that we do not actually binary-search towards a threshold between two cases here; we merely attempt to reach an overflow/underflow case.  It is not necessary to find precise threshold points to learn the rough magnitude and sign of the inputs.

Recovering the signs for each weight is useful since some of the cases (most of the ``normal'' ones) are symmetrical about zero, meaning we cannot learn the sign of the input only based on the state.  Recovering the sign before we begin the binary search means we always know exactly where we are in the state space.  Searching ``up'' in this calibration step gives us a sensible starting point to search down and thus reduces the number of invocations needed in the next phase.

\subsubsection{The Binary Search Phase}
Armed with the information collected in the calibration phase, we continue the attack against the same neuron as in phase 1 by finding the desired number of convergence point sets.  For as many convergence sets as we need, we start by generating an input set that is made of $n$ locked random values (one for each input) except that the $i$'th input is instead set to the large positive value recovered during calibration.  We feed the input into the network and note the return case by interpreting the CPU instruction counts trace for the neuron.  Even if some of the weights have different signs, the overall $\Sigma$ should be dominated by the large input multiplied by its weight.  If the neuron has overflowed, our input has caused the neuron to enter the state above the threshold between normal (+) and overflow and should search down.  If the neuron has instead underflowed, we must be in the boundary between underflow and normal (-). If the activation's input is negated before being passed into \expfunc{} (as in \sigmoidfunc{} ), the direction to search should be reversed. We continue the binary search on that $i$'th input, searching back up if we fall below an underflow/overflow case in terms of magnitude.  Eventually the search ends either because the desired search depth was reached or because we reached the limit of precision for the underlying floating point maths.

Using this algorithm, we can generate as many novel convergence sets as we require since the unchanging members of the input set are all randomised between inputs.  It is worth noting that the system of equations does not produce usable results if the values on the right hand side are all the same threshold, so targeting multiple thresholds is required.  One quick way to find more thresholds for the system of equations is to flip the signs of the inputs to ensure both the underflow-to-normal and overflow-to-normal thresholds are included in the solution set.  Note however that it is not required to target the thresholds between underflow/normal and normal/overflow; these are just the easiest to find since we can force an under/overflow and then work our way up/down until we find the normal region.  There are other thresholds within \expfunc{} (as shown in \Cref{fig:expf_state_map}) that can be exploited if they can be reached so long as we know the sign of the weight and thus in which precise state we are in.

Once we have the requisite number of equations, we can solve the system of equations as described in \Cref{sec:attack-concept} to recover the weights/bias for this first neuron.  We then perform the exact same two-part algorithm on each remaining neuron until we solve the entire layer.  The output from our recovery tool as it recovers one neuron is depicted in \Cref{sec:neuron_centric_attack}.

\subsection{Extending the Attack to Deeper Layers}\label{sec:deeper-layers}

In order to extend the attack into subsequent layers, we must solve two key problems.  First, how can we mitigate or `unwrap' already-solved early layers to insert arbitrary inputs beyond them, deeper into the network?  And second, is it always possible to generate those values?

\mypar{Unwrapping Solved Layers} Our attack methodology for the first layer relies on being able to precisely craft a set of ``target'' in order to search towards threshold points.  In order to be able to do this for deeper layers, we first have to ``unwrap'' all the recovered layers between the input layer and the layer we target.  In \Cref{sec:layer_unwrap}, we show the simple linear algebra required to rearrange the normal operation of a neuron to isolate the input as opposed to the target.

In these equations, $act$ and $actinv$ are known since we know the architecture of the network and $W$ and $b$ are known because they belong to a solved layer.  Therefore, given these recovered parameters, a desired `target' set of values, and the appropriate inverse activation function, we can calculate the input we must feed into the layer to produce a desired target as the output of that layer.  Note this assumes the activation function is invertible without a substantial loss of accuracy; inverting these functions might lead to an inability to pass certain values forward, which we discuss below.

\mypar{Expressive Power}
Even if we can algebraically calculate how to pass arbitrary values through a solved layer, subsequent layers are harder to attack because we are limited by the ``expressive power'' (see \Cref{tab:function_vuln}) of the neurons before the target layer in terms of both magnitude and sign.  There are two consequences to this.

The first is some values may be impossible to generate in subsequent layers.  For example, since the output of \sigmoidfunc{} is always positive, baring negative biases, it would be impossible to send negative values into the next layer.  Similarly, \relufunc{} can never produce a negative output.  The second is that certain thresholds such as those between \expffunc{}'s overflow/underflow and normal cases have a high magnitude ($\approx 88.723{-}103.972$) relative to generally smaller magnitude weights and biases.  \sigmoidfunc{} has a very small output range ($0{-}1$) making it impossible to produce a large enough $\Sigma$ in the next layer to `reach' one of these two thresholds unless either the network is very dense with small-magnitude positively weighted neurons or has some positive and large magnitude ($>1$) weights connected to it. This suggests denser networks are more expressive and therefore more vulnerable. In both of these cases, we rely on certain network properties.  Though this means that certain networks might not be vulnerable to this method, security guarantees should never rest on architectural decisions.

\mypar{Attack Outline}
Nevertheless, extending the attack into deeper layers is possible with certain caveats.  There is no need to focus only on the high magnitude thresholds we previously discussed between the underflow/overflow cases and the normal one; there are six in the range from $\approx$ -1.04--1.04 (see \Cref{fig:expf_state_map}).  By searching anywhere within this region (\eg near the subnormal case as in \cite{Gongye20}) we can, as before, build convergence sets for our inputs and solve for the hidden parameters.
Because we might pass through a low expressiveness activation function like \sigmoidfunc{} (output range from 0--1), which is likely being further reduced by multiplication with a weight with magnitude <1, we might not cross a single threshold point (the presence of the bias as a component of $\Sigma$ means that we might be shifted away from zero far enough that we cannot reach both sides of the subnormal case).

As such, we instead start with a grid search to scan (at some resolution up to the attacker which trades off speed and the chances of finding a narrow instruction count class) for multiple instruction count classes.  As with the binary search, we start by creating an input set that is made up of $n$ locked random values (one for each input) except that the (randomly chosen) $i$'th input is nominated as a dynamic value, initially set to a high magnitude positive number.  We scan the dynamic value from its initial value down to a high magnitude negative number, we can check to see if encounter multiple return cases.  If we are able to reach multiple classes, we can then perform a binary search between those classes to isolate the threshold and build a convergence point set, as before.

\subsection{Estimating Attack Performance}\label{sec:estimating_attack_performance}

We note that the number of executions to perform our attack is dependent on the architecture of the network (and therefore the number of parameters to recover) but also some factors the attacker can control such as the depth of the binary search.  Naively, the total number $Q$ of queries to recover the first layer can be computed as: $Q = \,P\,\cdot\,N\,\cdot S$,
%
%
where $P$ is the number of hidden parameters (in the first layer, \eg one weight per input plus one for the bias) to solve for, $N$ is the number of neurons in the first layer, and $S$ is the maximum number of steps in the binary search, minimised by the calibration process described in \Cref{sec:calibration}.

The $P$ value correlates to the minimum number of equations we require to solve a system of equations: one for each unknown.  We note that we can compensate for noise or other errors by adding extra equations to produce better solutions at the cost of more queries.  In testing, finding 3 additional convergence sets helped reduce the maximum error by 28\% at depth 25 compared to an attack where precisely $P$ convergence sets were generated (this is visualised in \Cref{sec:extra_equations}).  It might be possible with future study to detect particularly weak systems of equations analytically and dynamically add extra convergence sets to improve the results.

\section{Experimental Evaluation}\label{sec:experimental_evaluation}

Here we introduce and share the results of our attack against three \ac{PoC} networks trained on different datasets in ``full'' Tensorflow (in Python) and then converted to \tfulite{} using the process described in \Cref{sec:collecting_traces}.

\subsection{\pocI: A Regression Model using \expffunc{}}\label{sec:pocI}

\mypar{Model}
We first focus on a regression model to predict home insurance costs, which was trained on the ``Medical Cost Personal Dataset''~\cite{choi18}.  The network has three hidden layers.  The first hidden layer contains 100 neurons that use the \code{Exponential()} activation function which is just a direct call to \expffunc{}.  There are 11 input nodes, meaning that there are in total 1200 parameters (1100 weights and 100 biases) that we target.  After the first layer, there is a second hidden layer of 10 neurons which uses a \relufunc{} based activation function.  A final layer contains a single \relufunc{} neuron.

\mypar{Recovery}
For this first network, we only focus on the first hidden layer.  Despite that, we note that:

\begin{enumerate}
	\item These choices give us the clearest method to discuss and explain the underlying vulnerability (secret-dependent control flow).
	\item More complex networks or types of networks do not inherently offer greater security unless the underlying problem  is addressed, even if demonstrating the attack might be more technically challenging.
	\item The attack is agnostic to the accuracy of the network.
\end{enumerate}

\mypar{Results}\label{sec:pocI_results}
Our attack fully recovered all the weights and biases for the neurons in the first layer of a \tfulite{} network running inside an SGX enclave.  After calibration, the full attack took 55 binary searches of the network per parameter to correctly recover each of the 1200 parameter in the first layer down to or beyond 3 decimal places of accuracy.  This is in contrast to \citeauthor{tramer_stealing_2016}'s budget of 100 queries per parameter.

During the attack, each first-layer neurons' parameters were recovered at around 99\% accuracy (when compared to the ground truth values) by <36\% of the way through the full search (around search depth 20).  The average error is <1\% at this point.  The error continues to slowly improve until the search hits the limits of a 32-bit floating point number at around depth 55.  These findings are further summarised in \Cref{sec:pocI_speed_accuracy}, which shows how the achieved average and maximum error in the recovered weights and biases varies with the number of queries per neuron in \pocI.

The maximum error rates were higher in our full attack on \pocI than we had seen in smaller scale testing.  We speculate this is due to numerical precision issues when solving certain neurons, particularly ones with very small weights or perhaps a combination of very low and very high magnitude weights.  During the binary search there may be an ideal magnitude for the non-dynamic input values based on the magnitude of the dynamic input.  We consider this an area for future study and discuss a workaround below.
As this network was substantially larger than those used in small scale testing, we developed two ideas to help accelerate the attack for the other case studies.

\mypar{Max Search Depth}
We note that the attack can be sped-up by only binary searching to a certain depth of iterations.  As can be seen in \Cref{sec:pocI_speed_accuracy}, we already achieve an < 1\% average error rate by depth 20.  Based on the desired level of accuracy, an attacker can trade off between time and fidelity to the original network and choose to only search to some depth, rather than letting the binary search finish.  We note that achieving sub 1\% precision may not be as necessary as it seems; the widespread adoption of quantisation has shown that models can still be quite performant despite the precision loss incurred by the lossy conversion of hidden parameters to 8 bit integer representations.  We also note that our sub 1\% precision attack has complexity below that of~\citeauthor{tramer_stealing_2016}, which also only targets classification networks and not continuous output networks like this one.

\mypar{Improved Method of Calibration}
As described in \Cref{sec:calibration}, the calibration phase saves us from wasting many search steps down from some large safe constant to the smallest values that will overflow/underflow the weights.  In the previous equation, $S$ is therefore optimised by starting the search from the smallest possible point that still overflows/underflows our target.  By automating this calibration process for each neuron, we ensure a search with the fewest steps possible.

Though we first implemented calibration as a per-neuron process (as previously described) we realised when trying to attack this model that this step could be performed across an entire layer of neurons simultaneously.  We can leverage the fact that our \sgxstep{} traces for a given input provide an instruction count measurement for all neurons simultaneously to perform a combined or input-centric calibration.  We first create an input array for the network that is all zeros (one for each input).  Next, we iterate through each index in the input, replacing the 0 with increasing powers of some constant, \eg 10 and passing the array into the network.  We then record when an input array has underflowed/overflowed all the neurons in the layer.   This maximal value of the power of that constant, $D$, when multiplied by $P-1$ (the number of inputs) allows us to calculate $C$, the number of executions required to calibrate the whole network:
%
$C = \,(P-1)\,\cdot D$.

Note that $C$ is independent of $N$ and $S$.  As $Q$ would likely be dominated by $N$ in a real network, this calibration phase should account for only a small fraction of the total queries in the full attack.  By adding $C$ to $Q$ we get a good approximation for the number of executions required to recover the first layer and calibrate faster than via the original method.

\subsection{\pocII: A Multi-Layer Model using \sigmoidfunc{}}\label{sec:pocII}

Next, we show how to attack a more realistic activation function and recover partial information from deeper layers.

\mypar{Model}
We now focus on a network trained to multiply two numbers with the following architecture: there are two inputs, followed by the first hidden layer which contains 4 \sigmoidfunc{} neurons.  Then there is a second hidden layer containing 8 \sigmoidfunc{} neurons followed finally by the final layer with a single \relufunc{} neuron.

\mypar{Recovery}
We follow the steps described in \Cref{sec:first_layer} and \Cref{sec:collecting_traces} to recover the first layer.  The only difference is that we exploit a different CPU instruction count leakage; instead of targeting \code{tlibc}'s  exponential function, we now attack the \tflite{} framework-level \sigmoidfunc{} function as described in \Cref{sec:framework_functions}.  This shows how our attack is adaptable to different and more realistic activation functions, as well as side channel leakages elsewhere in the \ac{ML} inference pipeline.  We then use the procedure described in \Cref{sec:deeper-layers} to attack the second layer.

\mypar{Results}\label{sec:pocII_results} This recovery is summarised in \Cref{sec:pocII_speed_accuracy} and is comparable to the results of our attack against \pocI.
We are also able to recover partial (signless) convergence sets for the neurons in the second layer.  For example, we can recover 35 sets for the first neuron in the second layer.  Given its 4 inputs and single bias term (5 total unknowns) these are many more convergence point sets than we would need for a solution.  However, because we do not have sign information, we are unable to conclusively solve for the neurons' parameters.

\subsection{\pocIII: MNIST with \sigmoidfunc{}}\label{sec:pocIII}

\mypar{Model} Lastly we discuss a more complex model which was trained on the MNIST dataset \cite{deng2012mnist}.  Our MNIST case study is much more complex than the others before it featuring $28 x 28 = 784$ inputs and a first layer with $128$ neurons.  Adding the biases, this network has $100,480$ hidden parameters in the first layer making it two orders of magnitude larger than \pocI.  For this study, we focus on only the first layer.

\mypar{Recovery} Here we combine parts of the recovery strategies of the first two case studies: a larger network which we attack using the framework-level \sigmoidfunc{} vulnerability described in \Cref{sec:framework_functions}.   As in the first model, we only attempt recovery of the first layer.

\mypar{Results}\label{sec:pocIII_results}
Because of the greater number of hidden parameters for this model, recovery using the methods described above would take millions of iterations and while feasible, not entirely practical.  In \Cref{sec:pocI_results} we discussed how the value $Q$ could be reduced by reducing the search depth, $S$.  Another way to reduce $Q$ is to try to minimise the $\,P\,\cdot\,N\, = 785 \cdot 128 = 100480$ term.  This is possible if we approach the binary search a different way, inspired by the input-centric calibration described previously.

Rather than a neuron-centric regime (focusing on each neuron one at a time and ignoring the effect of the changing inputs on the other neurons in the same layer), we instead record the state of all the neurons in the layer simultaneously as they respond to changes to each input of the network in sequence.  We note that in this model's first layer, the weights are normally distributed meaning that there are clusters that can be easily searched together.  This allows us to exploit the structure of the distribution of the hidden parameters, group weights with similar magnitudes, and combine searches allowing better than $\,P\,\cdot\,N\,$ performance.  This also allows us to abandon searches in regions without any neurons.  See the output of this process over an entire input in \Cref{sec:input_centric_attack} and a visual in \Cref{sec:search_strategy_comparison}.  Note that this approach also integrates the input-centric calibration discussed above, which is completed for this input in 18 executions of the network.  Also note that every iteration before depth 18 (where each neuron is effectively in its own search space) is a savings against the more naive approach described above.

The output in \Cref{sec:input_centric_attack} recurses to a minimum gap size of only $0.1$, so it does not have the same precision as the previous attack code example from \Cref{sec:neuron_centric_attack}.  Note that the output of the input-centric attack to depth $0.1$ (1050 executions) is roughly analogous in recovery precision to a depth $20$ neuron-centric attack ($128 * 20 = 2500$ executions) meaning that we recover the same value using 42\% of the executions.  Extrapolating the full attack to maximum depth, we project that the neuron-centric approach would take $2,460,800$ executions (without calibration) and this input-centric approach would take $1,180,765$ executions with calibration, or 48\%.

The \citeauthor{tramer_stealing_2016} attack requires a budget of 100 times the number of parameters in the network.  If we only consider the first layer, this network would require a budget of $ (28 * 28 + 1) * 128 * 100 = 10,048,000 $ queries (there are 28*28 weights between each input and each neuron, and we add one extra parameter for the bias term).  Using the values above, we project a full depth scan of this layer in  $1,180,765$ queries, which is 11\% of the \citeauthor{tramer_stealing_2016} budget.

While we cannot report the accuracy of the full recovery, spot checks against ground truth values were very promising and could be made more (or less) precise by tuning the min gap parameter to stop the search when desired.



\section{Ecosystem Analysis}\label{sec:ecosystem_analysis}

We examined implementations of common activation functions in popular \ac{ML} frameworks to assess if they exhibit input-dependent patterns. For this, we used a combination of static (Ghidra) and dynamic (gdb) analysis to explore the call stacks of these frameworks until an underlying maths function was reached.  Our findings are summarised in \Cref{tab:library_survey}.

There are several dimensions to our survey: first, we found that in all cases we tested, the low-level maths functions that underpin activation functions are not included in the \ac{ML} frameworks themselves.  Instead, they rely either on additional high performance maths libraries, \eg \code{sleef} in the case of PyTorch, \code{Eigen} and \code{DNNL} in the case of Tensorflow and \code{XNN} in the case of Tensorflow Lite or call into standard libraries like \code{tlibc} (in the case of \ac{SGX}) or \code{glibc} on a Linux system.  In some cases, these frameworks do both for different functions, adding to the complexity of the analysis.

The (in)security of an activation function can stem from the underlying library and implementation of these functions.  For example, a call to \code{sigmoid()} that ends up in \code{glibc}'s highly input-dependent \expfunc{} in \code{libm.so} exposes the caller's ML framework to vulnerability.  Similarly, a call to the same function in a different framework that ends up in a highly optimised maths library like \code{Eigen} may be harder to exploit.  However, as we note in \Cref{sec:framework_functions} and \Cref{sec:pocII}, higher level framework code can also be the culprit of exploitable memory access patterns.  The operational context may also play a roll in vulnerability; e.g., an embedded edge device may not have the hardware functionality for high performance or even floating point maths routines.

The reason optimised maths libraries may be harder to exploit is their use of vectorised instructions (\eg {AVX} on Intel or {NEON} on \ac{ARM}) which discourage branching (since that can affect performance).  For example, we observed that in PyTorch, \code{softmax()} was realised through calls into the \code{sleef} library, which at default optimisations compiled into AVX assembly code without memory access leakage. We note however that \code{sleef} incorporates a \emph{dispatch} mechanism that calls into different implementations depending on the outcome of \code{cpuid}. Notably, if this library was used in an \ac{SGX} enclave, where \code{cpuid} is typically implemented as an \emph{untrusted }\code{ocall}, the adversary could spoof the \code{cpuid} result and force usage of the insecure non-AVX version.  Curiously, for \code{sigmoid()}, PyTorch resorts to the standard library implementation of \expfunc{}, rendering the implementation insecure independent of \code{cpuid}.  Tensorflow uses \code{Eigen} for all the functions we tested except for \code{Exponential()}, which calls directly into \code{glibc}.  Finally it is worth restating that even though vectorised code is non-branching, it is not guaranteed to execute in constant time~\cite{Gongye20}.  

Finally, \Cref{tab:library_survey} shows how the optimisation level has a direct impact on the security of \relufunc{} for \tfulite{}. While the compiler default of \code{-Os} happens to be free of memory access leakage, this is more by accident than by design, which is not desirable in the case of libraries that handle sensitive data.  Analogously, the security benefits of vectorised code against our attack is not a proactive security-related decision but a side effect of pursuing performance.

\begin{table}[tbp]
	\caption{Survey of vulnerability across various \ac{ML} and standard libraries and for different optimisation levels.}
	\label{tab:library_survey}
	\centering
	\small
	\setlength\tabcolsep{2pt}
	\resizebox{\columnwidth}{!}{
		\begin{tabular}{llcllll}
			\toprule
			\bf \ac{ML} Framework               & \bf RELU & \bf Exponential & \bf Sigmoid & \bf Softmax \\
			\midrule
			TFLiteMicro (SGX/tlibc, \code{-O0}) & \xmark   & \xmark          & \xmark      & \xmark      \\
			TFLiteMicro (SGX/tlibc, \code{-Os}) & \cmark   & \xmark          & \xmark      & \xmark      \\
			\midrule
			TFLiteMicro (\code{glibc}, \code{-Os})		& \cmark   & \xmark          & \xmark      & \xmark      \\
			TFLite{} (\code{glibc})                      & \cmark   & \xmark          & \cmark      & \cmark      \\
			TensorFlow CPU (\code{glibc})              & \cmark   & \cmark          & \cmark      & \cmark      \\
			PyTorch (\code{glibc})                     & \cmark   & \na             & \xmark      & \cmark*     \\
			\bottomrule
		\end{tabular}
	}
	\\ \smallskip
	\scriptsize
	Legend:
    \cmark: Secure
    \xmark: Insecure due to data-dependent access pattern\\
	*: Only secure if \code{cpuid} indicates AVX support
    \vspace{-6.5mm}
\end{table}

\mypar{Vulnerability of \relufunc{}}\label{sec:relu}
In addition to \expffunc{}, we also considered \relufunc{} and its implementation in \tfulite{}, which relies on the \code{std::max()} function.  When running inside of an enclave, the \code{std::max()} library call is provided by an implementation in \acp{SGX} \code{tlibc} library which can be seen in \Cref{sec:max_in_c}.  With the default settings (\code{gcc 11.4.0 -Os} with function marked \code{inline}), the above code was compiled into the instructions shown in \Cref{sec:max_in_asm}.

When compiling with \code{-O0}, however, the function was assembled as shown in \Cref{sec:max_in_asm_o0}.  This is significant because when the compiler emits \code{jmp}-class instructions, it is possible (using \code{sgx-gdb}) to detect the single step difference and therefore discern between the cases where $ input \leq 0 $ and $ input > 0 $.  This is a threshold point between two instruction count classes about $0.0$, similar to the ones discussed previously in \expffunc{}.  Though the difference might only be a single CPU instruction (whether a jump was taken or not) this difference is detectable.  

\Cref{sec:max_in_asm}, on the hand, has no step count differences due to its use of the \code{CMOVA} instruction which combines a branch and a move into a single instruction.  Whatever the outcome of the conditional, \code{CMOVA} always takes the same number of steps, removing the timing leak.

Within our case study, \relufunc{} within \tfulite{} was not vulnerable, but it was not secure by design either.  This is worrisome because in contrast to vetted cryptographic code, ML libraries currently \emph{do not} exhibit input-independent memory access and execution behaviour, and further the eventual security properties of high-level \ac{ML} code may depend on compiler optimisations and other non-explicit behaviour as with \relufunc{} in some configurations.  Without any specific safeguards taken, \ac{ML} libraries are not intrinsically secured against the vulnerability discussed in this paper.

We note that on x86, many floating point instructions can raise exceptions or otherwise exhibit operand-dependent timing as mentioned by \citeauthor{Gongye20} and (in a different context) \citeauthor{Kohlbrenner17}, which might result in timing vulnerabilities even if branchless code is emitted \cite{Gongye20,Kohlbrenner17}. We leave the exact measurement and exploitation of such timing side channels in real-world settings as an area for future research.

\mypar{Vulnerability of Framework Code}\label{sec:framework_code_vuln}
As discussed in \Cref{sec:framework_functions}, and demonstrated practically in \Cref{sec:deeper-layers}, exploitable memory access pattern vulnerabilities can exist outside mathematical library code in \ac{ML} frameworks themselves.  This further supports our case that these frameworks are not suitable for enclave execution under a \ac{SCA} advisory.  Beyond the security properties of the libraries they use, security-conscious \ac{ML} framework developers need to also consider their own code and if they are introducing exploitable timing side channels.  This is especially true for more complex activation functions like \code{softmax()} which are not mathematical functions likely to be found in standard maths libraries.

\section{Discussion and Mitigations}\label{sec:discussion}

We show that input-dependent memory accesses and branching are prevalent in \ac{ML} implementations, and that the necessity for secure programming practices is not widely understood in the \ac{ML} community.  Our \acp{PoC} shows that an attacker with a high-resolution controlled or side channel is able to exploit data-dependent memory accesses to recover model details, whereas prior software \ac{SCA} typically only recovered hyperparameters such as model architecture~\cite{yan_cache_2020}. Though there are certain limitations, \pocI and \pocII show precise weight recovery can be performed with at maximum 55 queries per weight, and at over 99\% accuracy at less than half that number (plus calibration).  Search depth can be parametrically adjusted to account for the desired accuracy.

Ultimately, while out of scope for this paper, we believe that if an attacker can obtain precise-enough timing measurements,  similar attacks may be possible outside of a \ac{TEE} context, \eg for virtualised cloud deployments. Cache attacks with high temporal resolution, \eg Prime+Scope~\cite{Purnal21}, could be enough to recover traces similar to \sgxstep{}, though attacks would likely require modifications to deal with increased noise.

Regardless of whether ML is being performed in enclaves, we argue that hence, \ac{ML} library developers should pay attention to the full implementation attack surface including library and framework code. Even if \eg \relufunc{} implementations were often secure in the surveyed libraries, this largely relied on certain compiler optimisations and was not `by design'. Similar to cryptographic libraries, \ac{ML} libraries should consider hardening their codebase against software side channels such as data-dependent memory access patterns.  Security and privacy-sensitive algorithms, to which \ac{ML} now belongs, must be robust by design, regardless of context, platform, or (mis)use of library interfaces by a programmer.

We briefly explored quantised networks as an extension to our work. With fewer bits representing each hidden parameter, we reasoned binary searches might encounter discrete boundaries in fewer iterations.  We converted our MNIST model using Tensorflow's full integer quantisation to a \tflite{} model with 8bit integer representations of its weights and biases and captured traces using our \sgxstep{} setup (an inset of which is visualised in \Cref{sec:quantised_trace}).  Aside from the additional (de)quantisation phases during the network's execution (not pictured), the patterns of page accesses and timings were very different to previous traces, owning to the different functions / processes used to perform inference on fixed point values.  Though we did not further investigate the access patterns for this paper, we noticed evidence of non-constant time behaviour in the evaluation of neighbouring neurons suggesting non-constant time effects.

\mypar{Countermeasures} As with the work done in cryptography libraries, there are established steps to avoid data-dependent branching and memory accesses, \eg by relying on bitwise logical operations, constant-time conditional moves, and other related techniques.  For x86, as discussed for the \code{sleef} library, using vectorised SIMD code can also mitigate the issue, though needs to be thoroughly vetted for other side channels like floating point timing~\cite{Kohlbrenner17,Gongye20}.

At present, akin to a ``chosen plaintext'' attack for cryptographic algorithms, our attacks entails sending arbitrarily large or small floating point numbers into the network. Any input normalisation might mitigate the fast calibration and binary search approaches described in this paper, only enabling partial recovery through the grid search technique.  Similarly, as the attack requires a large number of queries to the network, rate-limiting or detection of ``suspicious'' usage would render exploitation harder in practice. However, again, \ac{ML} security should in our view not rest on developers having to normalise inputs or limit performance, so these measures are supplementary to a securely developed library.

Finally, researchers have explored defenses at the level of the TEE itself to frustrate (but not fully eliminate) side-channel leakage exploitation.
Initial approaches~\cite{shih2017tsgx,gridin2020dejavu,oleksenko2018varys} focused on detecting suspicious interrupt rates as a side-effect of an ongoing controlled-channel attack, which may be error-prone and suffer from false positives.
Alternatively, custom oblivious RAM solutions~\cite{aga2019invisipage,zhang2020klotski,brasser2019dr} may probabilistically hide secret-dependent enclave memory accesses, however this is at the cost of prohibitive performance overheads.
Most prominently, AEX-Notify~\cite{constable2023aexnotify} is a recent Intel SGX hardware-software extension that aims to eliminate \sgxstep{}'s single-stepping capabilities by prefetching selected application pages.
Notably, AEX-Notify is an \emph{opt-in} feature only available in recent CPUs.
Enclaves that did opt-in would preclude the ability to gather precise instruction counts through single-stepping. However, AEX-Notify explicitly does \emph{not} prevent general information leakage through page faults and, hence, cannot fully mitigate our attacks.
While specialised, compiler-assisted defenses~\cite{shinde2016preventing,vanoverloop2025tlblur} have been proposed to mitigate page-fault leakage on off-the-shelf SGX platforms, these approaches incur prohibitive performance overheads, and their applicability to general-purpose ML workloads remains unexplored.

\mypar{Limitations} We require the victim to run in a co-located TEE in order to use our controlled channel, which is a different requirement than previous model stealing attacks, that require chosen inputs and the output distribution of the inference pass.

Despite the recovery of the first layer of neurons fundamentally undermining SGX’s promise to secure the entire computation, our attack is limited in recovering parameters from deeper layers, depending on the model architecture. This is mostly due to the (limited) expressive power of some common activation functions as noted in \Cref{tab:function_vuln,sec:deeper-layers}. The grid search is less efficient than the binary search used for the first layers, and depending on the magnitude of the weights, small search steps might be needed to discover thresholds.
An adaptive approach could deepen the sweep region until a threshold is hit; however, small weights likely contribute little to the network and may be replaced by a static value to produce an equivalent (non-identical) network.

Finally, we focused on input-dependent memory access patterns and instruction counts in this paper, but did not take input-dependent latency of single instructions into account.  For example, floating point multiplication is not necessarily guaranteed to execute in constant time~\cite{Gongye20}.  This could render vectorised code like the implementation from \code{sleef} vulnerable to a modified version of our attack: while different inputs might have identical instruction counts (as reported by \sgxstep{}), the actual runtime might be different and be observable with other side channels like interrupt latency~\cite{VanBulck18}.

\section{Conclusion}

The recent surge in ML applications necessitates an in-depth understanding of new attack vectors and defenses.  In this respect, confidential computing architectures, and in particular Intel SGX, have gained significant traction in recent years and hold the potential to securely outsource critical ML computations to pervasive untrusted remote cloud platforms.  However, we show that SGX-protected ML implementations exhibit subtle side-channel vulnerabilities that allow the extraction of precise model parameters in our \ac{PoC} attacks.
In the wider perspective, our work may generalise to other \acp{TEE} and highlights the issue of input-dependent memory accesses that are prevalent in today's ML implementations, suggesting that security-critical ML applications should consider adopting coding practices studied in the cryptographic community.

\section*{Acknowledgments}
The PhD of Jesse Spielman is partially funded by a donation from Intel. This research was further partially funded by the Engineering and Physical Sciences Research Council (EPSRC) under grants EP/X03738X/1 and EP/R012598/1.
This research is partially funded by the Research Fund KU Leuven, and by the Cybersecurity Research Program Flanders.


\printbibliography

\appendices

\begin{appendices}

\section{Implementation of \sigmoidfunc{} (called \code{logistic()}) in \tflite{}}\label{sec:sigmoid_in_tflite}
\begin{figure}[H]
\begin{lstlisting}[label={lst:sigmoid_in_tflite}, caption={Implementation of \code{Logistic()} in \tflite{}}, captionpos=b, escapechar=§, linewidth=0.95\columnwidth]
inline void Logistic(const RuntimeShape& input_shape, const float* input_data,
                     const RuntimeShape& output_shape, float* output_data) {
  const float cutoff_upper = 16.619047164916992188f;
  const float cutoff_lower = -9.f;

  const int flat_size = MatchingFlatSize(input_shape, output_shape);

  // [comments removed]

  for (int i = 0; i < flat_size; i++) {
    float val = input_data[i];
    float result;
    if (val > cutoff_upper) {
      result = 1.0f;
    } else if (val < cutoff_lower) {
      result = std::exp(val);
    } else {
      result = 1.f / (1.f + std::exp(-val));
    }
    output_data[i] = result;
  }
}
\end{lstlisting}
\end{figure}

\FloatBarrier

\section{\expffunc{} call stack for \tfulite{} neuron}
\label{sec:sgx_expf_stack}
\begin{figure}[H]
\begin{lstlisting}[label={lst:sgx_expf_stack}, caption=Simplified call stack for \expffunc{}-based neuron in \\ \tfulite{}., captionpos=b, escapechar=§, linewidth=0.95\columnwidth]
0x7ffff762b460 in expf()
0x7ffff7625e9c in tflite::(anonymous namespace)::ExpEval()
0x7ffff762318b in tflite::MicroInterpreterGraph::InvokeSubgraph()
0x7ffff760f04e in performInference()
0x7ffff7610577 in entry_point()
0x7ffff7611633 in sgx_entry_point()
0x7ffff76130a2 in do_ecall()
0x7ffff762c5a5 in enter_enclave()
0x7ffff762c7d5 in enclave_entry()
0x7ffff7f95309 in __morestack()
0x7ffff7f986fb in CEnclave::ecall()
0x7ffff7f9a852 in _sgx_ecall()
0x55555555bac9 in entry_point()
0x555555557a24 in main()
\end{lstlisting}
\end{figure}

\FloatBarrier

\section{Neuron-Centric Attack Output}\label{sec:neuron_centric_attack}
\begin{figure}[H]
\begin{lstlisting}[label={lst:neuron_centric_attack}, caption=Example output from our neuron-centric attack tool \\ finding convergence points., captionpos=b, escapechar=§, linewidth=0.95\columnwidth]
	[ Neuron 2 ]
	stats:
		952.99 seconds since start
		1439 executions so far
		4 Incomplete logs
		0 non-deterministic steps
		6 neuron_check failures
	Function = exp
	Search strategy: seeded binary search
		Calibrating:
			count: 14
			recovered signs: + - - - + - - - - + +
			recovered maxvals: 1000 1000 1000 1000 1000 10000 1000 1000 10000 10000 1000
		Finding convergence points for equation 1
			depth 0: Overflow
			depth 10: Normal
			depth 20: Overflow
			depth 30: Normal
			depth 40: Overflow
			depth 50: Overflow
		Finding convergence points for equation 2
			<SNIP>
		Finding convergence points for equation 12
			depth 0: Overflow
			depth 10: Normal
			depth 20: Normal
			depth 30: Normal
			depth 40: Overflow
			depth 50: Overflow
	deepest solution: [
		0.14825567 -0.17573831 -0.28457861 -0.24754734  
		0.09355622 -0.09359772 -0.2866397  -0.2083804  
		-0.10068617  0.07542896  0.10219476 -0.04099777]
	ground truth: [
		0.14825566 -0.17573832 -0.28457862 -0.24754737  
		0.09355621 -0.09359772 -0.28663969 -0.2083804  
		-0.10068618  0.07542896  0.10219476 -0.04099637]
	abs percent error: [
		0.0%, 0.0%, 0.0%, 0.0%, 0.0%, 0.0%, 
		0.0%, 0.0%, 0.0%, 0.0%, 0.0%, 0.0%]
	Checkpoint saved!
\end{lstlisting}
\end{figure}

\FloatBarrier

\section{Layer Unwrapping}\label{sec:layer_unwrap}
\begin{figure}[H]
\begin{align*}
	act(input * W + b) &= target \\
	input * W + b &= act^{-1}(target) \\ 
	input * W &= act^{-1}(target) - b \\
	input * \cancel{W * W^{-1}} &= (act^{-1}(target) - b) * W^{-1} \\ 
	input &= (act^{-1}(target) - b) * W^{-1} \\
\end{align*}
\caption{Note that it is also possible to do this without matrix inversion for greater numerical efficiency, \eg using \code{numpy.linalg.lstsq(W, act\_inv(target)-b)}.}
\end{figure}

\FloatBarrier

\section{\pocI Results}\label{sec:pocI_speed_accuracy}
\begin{table}[H]
	\caption{Average and Maximum Error for different search \\ iterations of \pocI}
	\label{tab:pocI_speed_accuracy}
	\small
	\setlength\tabcolsep{2pt}
        \begin{tabular}{c S[table-format=3.15] S[table-format=3.13]}
            \toprule
            \bf {Iterations} & \bf {Average error}        & \bf {Max.\ error}        \\
            \toprule
                 5		      & 6.029863100281893 & 844.0875066255469 \\
                 10  	      & 2.836460066697138 & 786.2672913954896 \\
                 15  	      & 0.271652801976045 & 156.0382954397002 \\
                 20  	      & 0.008040720460419 &   4.5014965290057 \\
                 25  	      & 0.000536511600745 &   0.2550122973594 \\
                 30  	      & 0.000389102878050 &   0.1506828257822 \\
                 35  	      & 0.000382396333823 &   0.1465018434229 \\
                 40  	      & 0.000382652272853 &   0.1466460219544 \\
                 45  	      & 0.000382651560351 &   0.1466454696212 \\
                 50  	      & 0.000382651519816 &   0.1466455095500 \\
                 55  	      & 0.000422195967951 &   0.1466455084049 \\
            \bottomrule
		\end{tabular}
	\\ \smallskip
	\scriptsize
\end{table}

\FloatBarrier

\section{\pocII Results}\label{sec:pocII_speed_accuracy}
\begin{table}[H]
    \caption{Average and Maximum Error for different search \\ iterations of \pocII}
	\label{tab:pocII_speed_accuracy}
	\small
	\setlength\tabcolsep{2pt}
        \begin{tabular}{c S[table-format=1.15] S[table-format=2.14]}
            \toprule
            \bf {Iterations}  & \bf {Average error}        & \bf {Max.\ error}		\\
            \toprule
                5			  & 5.104921006404077 & 17.91686599947229 \\
                10  		  & 1.675645751550916 & 17.51883507815715 \\
                15  		  & 1.529534168402632 & 17.52095831363468 \\
                20  		  & 0.069626021904943 &  0.79887716427850 \\
                25  		  & 0.002053382484092 &  0.02355709795055 \\
                30  		  & 0.000041334563160 &  0.00045190244010 \\
                35  		  & 0.000033134719817 &  0.00036192843021 \\
                40  		  & 0.000032159960239 &  0.00035062330641 \\
                45  		  & 0.000032157624911 &  0.00035059400594 \\
                50  		  & 0.000032157548965 &  0.00035059308250 \\
                55  		  & 0.000032157569172 &  0.00035059331140 \\
            \bottomrule
		\end{tabular}
	\\ \smallskip
	\scriptsize
\end{table}

\FloatBarrier

\section{Input-Centric Attack Output}\label{sec:input_centric_attack}
\begin{figure}[H]
\begin{lstlisting}[label={lst:input_centric_attack}, caption=Example output from our optimzied input-centric attack tool finding convergence points.  Note how after iteration 18 each of the 128 neuron is in it's own gap., captionpos=b, escapechar=§, linewidth=0.95\columnwidth]
Equation 3 / 785 / Other Value: 0.017745 / Input idx: 2
Calibrating...
Calibrated in: 18 executions to 131072
iteration: 0 / gaps: 1 / spans: 131072.0
iteration: 1 / gaps: 2 / spans: 65536.0
iteration: 2 / gaps: 2 / spans: 32768.0
iteration: 3 / gaps: 2 / spans: 16384.0
iteration: 4 / gaps: 3 / spans: 8192.0
iteration: 5 / gaps: 5 / spans: 4096.0
iteration: 6 / gaps: 7 / spans: 2048.0
iteration: 7 / gaps: 9 / spans: 1024.0
iteration: 8 / gaps: 12 / spans: 512.0
iteration: 9 / gaps: 17 / spans: 256.0
iteration: 10 / gaps: 25 / spans: 128.0
iteration: 11 / gaps: 34 / spans: 64.0
iteration: 12 / gaps: 43 / spans: 32.0
iteration: 13 / gaps: 55 / spans: 16.0
iteration: 14 / gaps: 74 / spans: 8.0
iteration: 15 / gaps: 94 / spans: 4.0
iteration: 16 / gaps: 108 / spans: 2.0
iteration: 17 / gaps: 117 / spans: 1.0
iteration: 18 / gaps: 123 / spans: 0.5
iteration: 19 / gaps: 130 / spans: 0.25
iteration: 20 / gaps: 134 / spans: 0.125
Done with equation #3/785 in 1050 iterations!
Saving equation 3 to cache
\end{lstlisting}
\end{figure}

\FloatBarrier

\section{\code{tlibc}'s \code{std::max()}) in \tflite{}}\label{sec:max_in_c}
\begin{figure}[H]
\begin{lstlisting}[label={lst:max_in_c}, caption=\code{std::max()} source code from \code{sgxsdk/include/libcxx/algorithm}., captionpos=b, escapechar=§]
  template <class _T1>
  struct __less<_T1, _T1>{
      _LIBCPP_INLINE_VISIBILITY _LIBCPP_CONSTEXPR_AFTER_CXX11
      bool operator()(const _T1& __x, const _T1& __y) const {return __x < __y;}
  };
  
  template <class _Tp, class _Compare>
  inline _LIBCPP_INLINE_VISIBILITY _LIBCPP_CONSTEXPR_AFTER_CXX11
  const _Tp&
  max(const _Tp& __a, const _Tp& __b, _Compare __comp){
      return __comp(__a, __b) ? __b : __a;
  }
\end{lstlisting}
\end{figure}

\FloatBarrier

\section{\code{std::max()} at \code{-Os}}\label{sec:max_in_asm}
\begin{figure}[H]
\begin{lstlisting}[label={lst:max_in_asm}, caption=Resulting assembly for \code{std::max()} at \code{-Os} with gcc 11.4.0., captionpos=b, escapechar=§]
MOVSS    XMM0,dword ptr [param_2]
UCOMISS  XMM0,dword ptr [param_1]
MOV      RAX,param_1
CMOVA    RAX,param_2
RET
\end{lstlisting}
\end{figure}

\section{\code{std::max()} at \code{-O0}}\label{sec:max_in_asm_o0}
\begin{figure}[H]
\begin{lstlisting}[label={lst:max_in_asm_o0}, caption=Resulting assembly for \code{std::max()} at \code{-O0} with gcc 11.4.0., captionpos=b, escapechar=§]
PUSH    RBP
MOV     RBP,RSP
MOV     qword ptr [RBP + local_10],param_1
MOV     qword ptr [RBP + local_18],param_2
MOV     RAX,qword ptr [RBP + local_10]
MOVSS   XMM1,dword ptr [RAX]
MOV     RAX,qword ptr [RBP + local_18]
MOVSS   XMM0,dword ptr [RAX]
COMISS  XMM0,XMM1
JBE     001019f0
MOV     RAX,qword ptr [RBP + local_18]
JMP     001019f4
MOV     RAX,qword ptr [RBP + local_10]
POP     RBP
RET
\end{lstlisting}
\end{figure}

\begin{figure}[H]
    \section{Error Comparison with Extra Convergence Sets}\label{sec:extra_equations}
	\includegraphics[width=\linewidth]{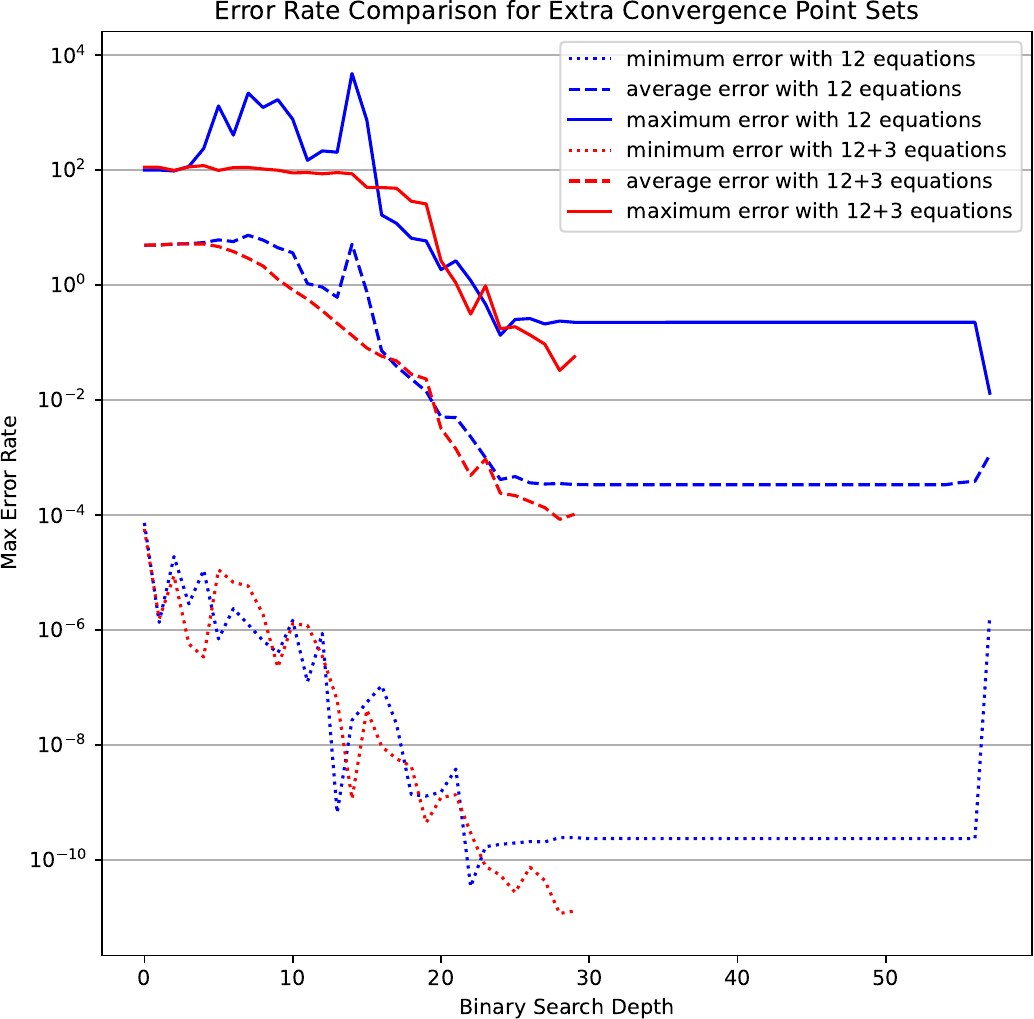}
    \caption{Note the effect that extra convergence sets have to reduce variance and lessen the errors, especially after depth=25}
	\label{fig:extra_equations}
\end{figure}


\begin{figure*}[tp]
    \section{Search Strategy Comparison}\label{sec:search_strategy_comparison}
	\includegraphics[width=\linewidth]{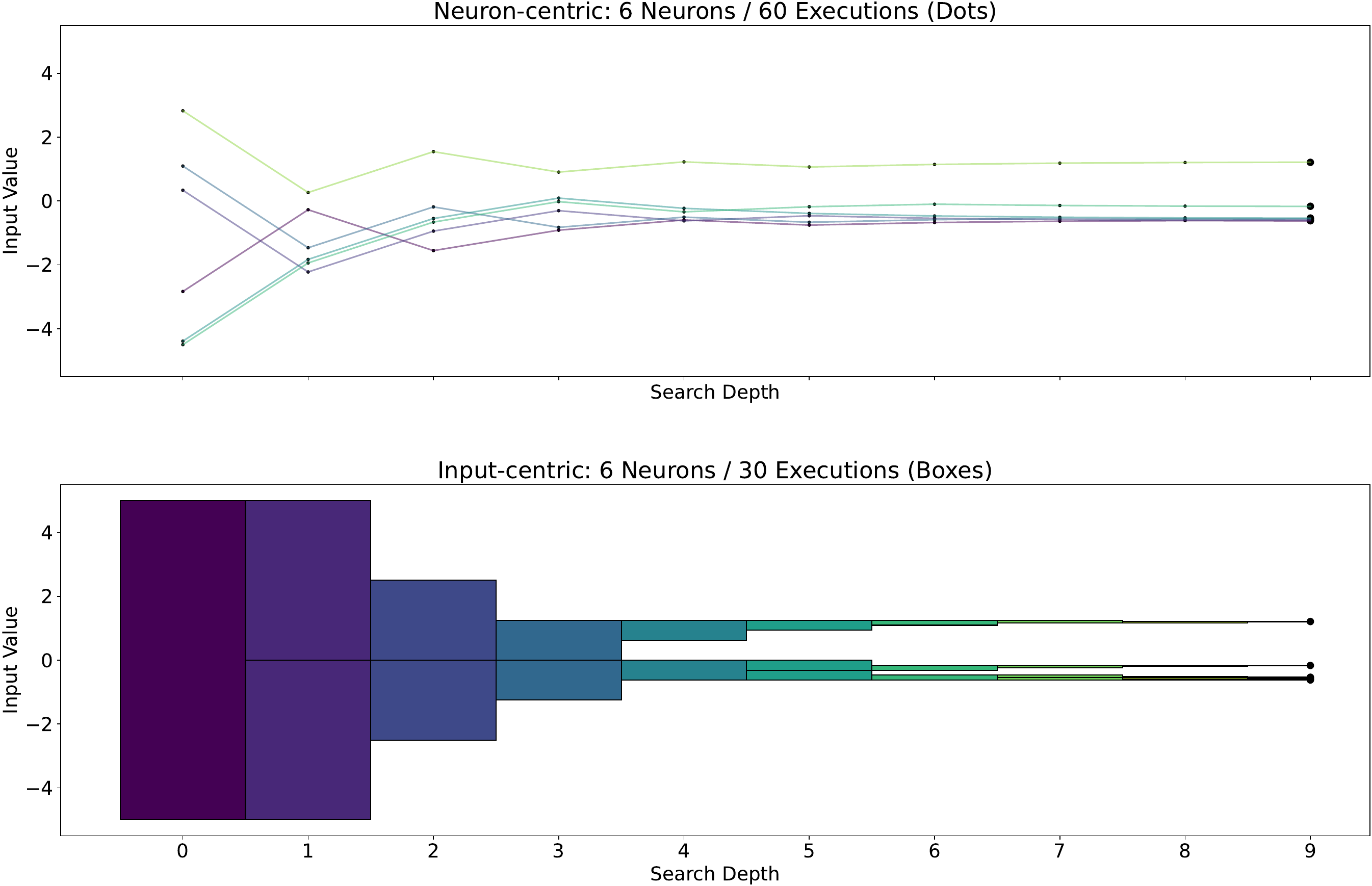}
	\caption{A visualisation comparing the execution count of a neuron-centric approach (top) and input-centric approach (bottom) for a single input and 6 normally distributed synthetic `neurons'.  For simplicity, we ignore the calibration process.  The neuron-centric approach is purely multiplicative (neurons times search depth) so for six neurons and a depth of 10 we cause 60 executions.  Here, each colour represents one neuron and the lines represent the binary search process.  \\ \\ The input-centric approach, on the other hand, exploits the distribution of the neurons to `save' executions by recursively bundling queries together for as long as possible until they have diverged into their own unique streams.  Due to the clustering of these 6 neurons, we can gain the same amount of information in half the number of executions as the first approach.  In this subplot, colour represents search depth.}
	\label{fig:search_strategy_comparison}
\end{figure*}

\begin{figure*}[tp]
    \section{Inset of trace from quantised MNIST network}\label{sec:quantised_trace}
	\includegraphics[width=\linewidth]{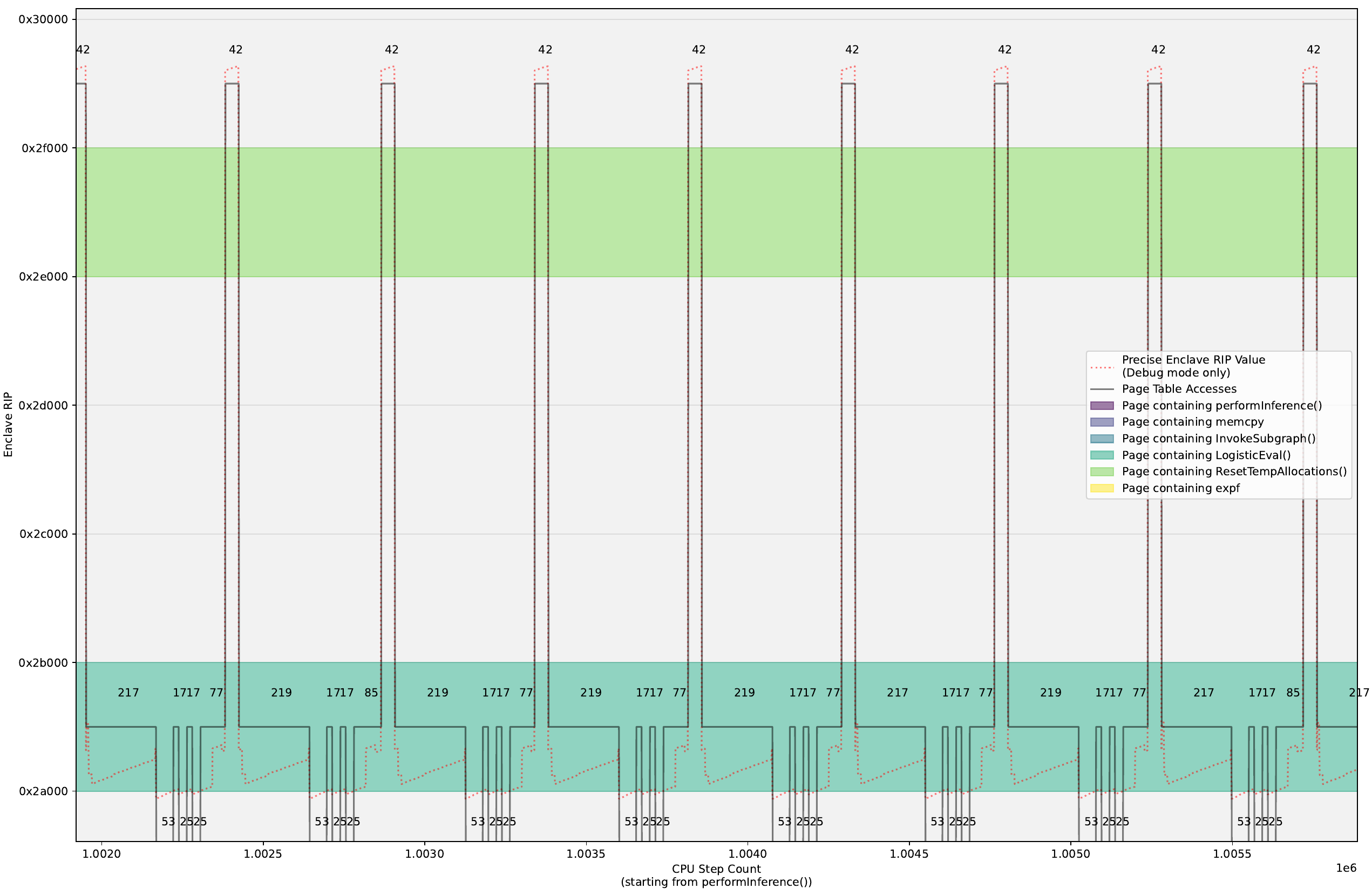}
	\caption{This is an inset of the region of the trace that corresponds to the 128 neurons in the first layer.  Note the largely but not entirely constant pattern of steps in the blue-green page at the bottom suggesting possible non-constant time behavior; the first and last values in each "valley" are either 217 or 219 and 77 or 85; everything else appears constant.  This difference between activations is echoed in the subtly different RIP patterns.}
	\label{fig:quantised_trace}
\end{figure*}

\end{appendices}
\end{document}